\documentclass[10pt,journal,  oneside, twocolumn]{IEEEtran}

\usepackage{multirow}
\usepackage{latexsym}
\usepackage{graphicx}
\usepackage{float}
\usepackage{amsmath}
\usepackage{amsthm}
\usepackage{lipsum}
\usepackage{subfig}
\usepackage{graphicx}
\usepackage{authblk}
\usepackage{bm}
\usepackage{amsthm}
\usepackage[section]{placeins}
\usepackage{wasysym}

\usepackage{colortbl}

\usepackage{enumitem}

\makeatletter  
\newif\if@restonecol  
\makeatother

\usepackage[linesnumbered,ruled,vlined]{algorithm2e}
\usepackage{algpseudocode}  
\usepackage{amsmath}

\usepackage{amssymb}

\usepackage{mathrsfs}
\usepackage{subfig}
\usepackage{caption}
\begin{document}

\title{RIS-aided Trajectory Optimization in Layered Urban Air Mobility}


\author{Kai Xiong,~\IEEEmembership{Member,~IEEE}, Supeng Leng,~\IEEEmembership{Member,~IEEE},  Liyuan Chen, Dapei Zhang, Chongwen Huang,~\IEEEmembership{Member,~IEEE}, Chau Yuen,~\IEEEmembership{Fellow,~IEEE}

\thanks{

K. Xiong, S. Leng, L. Chen, and D. Zhang are with School of Information and Communication Engineering, University of Electronic Science and Technology of China, Chengdu, 611731, China; and, Shenzhen Institute for Advanced Study, University of Electronic Science and Technology of China, Shenzhen, 518110, China.
}

\thanks{
C. Huang is with College of Information Science and Electronic Engineering, Zhejiang University, Hangzhou 310027, China, and Zhejiang Provincial Key Laboratory of Info. Proc., Commun. \& Netw. (IPCAN), Hangzhou 310027, China.
}

\thanks{
C. Yuen is with the School of Electrical and Electronic Engineering, Nanyang Technological University, Singapore. 
}






\thanks{The corresponding author is Supeng Leng, email: \{spleng, xiongkai\}@uestc.edu.cn}
}


\maketitle


\begin{abstract}
Urban Air Mobility (UAM) relies on developing aerospace industries, where safe aviation and efficient communication are critical features of aircraft.
However, it is challenging for aircraft to sustain efficient air-ground communication in urban circumstances. 
Without continuous air-ground communication, aircraft may experience course deviation and safety accidents.
To address these problems, a reconfigurable intelligent surface(RIS)-aided trajectory optimization scheme is proposed enabling efficient air-ground communication and safe aviation in UAM with a layered airspace structure.
This paper first devises a dual-plane RIS communication scheme for layered airspace. 
It fully engages the omnidirectional and directional signal attributes to reduce the transmission delay of the air-ground communication.
Based on the dual-plane RIS configuration, we jointly develop the intra- and inter-layer trajectory scheme to optimize communication and safe aviation.
In the intra-layer trajectory optimization, we propose a dual-time-scale flight scheme to improve communication capacity and horizontal flight safety.
Meanwhile, we propose a safe layer-switching method to ensure collision avoidance during vertical flight in the inter-layer trajectory optimization.
The communication load of the proposed scheme can be improved $40\%$ and the time of safe separation restoration can be lessened $66\%$ compared with the benchmarks in the layered airspace.



\end{abstract}


\IEEEpeerreviewmaketitle

\section{Introduction}

\IEEEPARstart {R}{oad} traffic congestion remains a significant hazard that causes severe damage to metropolitan daily life. 
Administrations aspire to enable aircraft technology as an alternative solution for modern transportation systems. 
The manned aircraft is envisioned to renovate existing ground and aerial transportation systems. 
An aircraft-based transportation system named Urban Air Mobility (UAM), proposed by the National Aeronautics and Space Administration (NASA), requires aircraft safe and efficient navigation in urban airspace \cite{Ehang3248}.  
Many institutions value the potential of UAM operations, such as Uber Elevate, Amazon Prime Air, and Unmanned Traffic Management \cite{Pinto9468356}.

To improve en-route airspace traffic safety and efficiency, a predefined airspace structure is required \cite{2015ComplexityAO}.
This airspace structure determines the separation of adjacent aircraft, ensuring collision avoidance and aviation efficiency.
Sunil \textit{et. al} \cite{42358726Sunil} validated that the vertically layered airspace structure can maximize the air traffic capacity underlying safe aviation.
The layered UAM can lessen traffic competition and maximize airspace utilization, which has been widely investigated \cite{2007Concept}.

Nevertheless, the en-route airspace is historically dependent on the coverage of radio beacons, dating back to the early 1950s \cite{Hoekstra345783278}.
This radio system was devised to help pilots navigate safely under ground-based Air Traffic Controller (ATCo) monitoring.
Hence, the aircraft trajectory design should fully consider the radio coverage and quality.
However, traditional radio communication meets new challenges for the specific layered airspace.
Typically, aircraft would like to enter a higher layer for fast travel. 
As a matter of fact, the available radio bandwidth declines with the flight altitude. 
The reason is that the base stations deployed for ground subscribers do not have sufficient communication coverage in the air.
This ground communication setup yields the contradiction between high-altitude shortage radio resources and travel demands \cite{{Zhou10246153},{Fan9638386},{Liu101049cmu212460},{Zhou9839387}}.


As an emerging 6th generation (6G) radio technology, reconfigurable intelligent surfaces (RIS) can manipulate radio waves to customize radio propagation features \cite{Xu9676676}.
Wherein, the ground base station communicates with high-layer aircraft through the RIS reflection mounted on low-layer aircraft or building surfaces.
Although involving RIS can address the airspace signal coverage and strength, it requires precise RIS phase shift control to align the reflected signal.
However, the velocity difference in the layered UAM attenuates the accuracy of signal alignment.
Typically, the flight velocity of aircraft rises with the layer altitude \cite{{42358726Sunil}}. 
For instance, short-travel aircraft can stay at low layers, while long-travel aircraft can fly at higher layers to improve travel efficiency.

{
Therefore, the developed trajectory optimization should consider both safe aviation and communication performance within the specific layered airspace.
However, due to the three-dimensional airspace, the proposed algorithm must simultaneously steer safe and communication-efficient flight at horizontal and vertical dimensions.
The horizontal flight is involved with intra-layer trajectory optimization. On the other hand, the vertical flight is involved with layer-switching optimization.



This paper proposes a RIS-aided trajectory optimization framework for safe and communication-efficiency aviation in layered UAM. 
A dual-plane RIS communication scheme is proposed to address the RIS alignment challenge.
Based on this RIS configuration, we devise a composite potential field to optimize the trajectory on the horizontal dimension.
Regarding the vertical dimension, we propose a layer-switching scheme inspired by the exponential back-off strategy that ensures collision-proof and separation recovery of adjacent aircraft.
Combining the above schemes, aircraft can travel safely and communicate efficiently in the horizontal and vertical dimensions.
The main contributions are summarized as follows:


\begin{itemize}
\item We proposed a dual-plane RIS communication scheme to perform low-latency communication in layered UAM.
The control plane utilizes the omnidirectional signal to convey the kinetic and control messages among the aircraft and the base station.
Leveraging these messages, RIS revises the phase shift for signal alignment.
This scheme can produce a low-latency transmission for aircraft in the high layer, with the RIS reflections of low-layer aircraft or building surfaces.
Simulation results exhibit that dual-plane RIS communication has the lowest delay upper bound and transmission failure probability.

\item We designed a novel horizontal flight control scheme for intra-layer trajectory optimization.
It comprehensively reconciles layered safe aviation with communication performance by large- and small-time-scale trajectory optimization.
The large-time-scale trajectory optimization derives the optimal relative positions between low- and high-layer aircraft to maximize the air-ground communication rate.
Conversely, the small-time-scale optimization produces five virtual potential fields to sustain safe navigation. There is a repulsive field for collision-free, an attractive field for predetermined travel, a stable field for flight stabilization, a goal field for approaching the small-time-scale optimal position, and a layer field to preserve the layered flight.  
Compared to traditional flight control, which only oversees flight kinetics, the proposed scheme takes communication into account.

\item We developed a collision avoidance layer-switching scheme on the vertical dimension for inter-layer flight optimization.
The layer-switching of an aircraft is triggered once the safe separation is violated in the course.
This layer-switching scheme refers to the design concept of the back-off algorithm.
the inter-layer airspace of the layered UAM is analogous to a communication channel, the layer-switching operation is analogous to message transmission, and the switching probability of aircraft is analogous to the back-off probability.
This scheme accelerates the safe separation restoration of vertical adjacent aircraft.

\end{itemize}
}
The remainder of this paper is organized as follows. Section II reviews the related works. Section III presents the whole UAM communication architecture.
The RIS-aided trajectory optimization scheme is presented in section IV. 
Section V provides the simulation results and the performance discussion. Finally, we conclude in Section VI. 

\section{Related Work}

The flight trajectory of aircraft highly relies on communication for safe aviation.
Nevertheless, ground base stations cannot yield acceptable communication coverage in the air \cite{Chen10149027}.
It is a research gap in efficient communication for layered UAM.
Although RIS is proposed as a promising new solution to unfold communication coverage in the air, it is inevitably impacted by the aircraft trajectory due to precise RIS alignment issues.
Hereafter, we retrospected UAM development, RIS communication, and archetype flight control investigations, respectively.

\subsection{Urban Air Mobility}
In 2021, the State Council of China issued the Outline of the National Comprehensive Vertical Transportation Plan, which proposed the Low-Altitude Airspace Economy for the first time. 
The low-altitude economy is driven by general aviation development, such as the flying car technology of aircraft. 
The flying car can run as regular vehicles on the road and fly as Vertical TakeOff and Landing (VTOL) aircraft in the air \cite{7420801Kaushik}.
Leading automobile companies like Audi, Toyota, Geely, and Xiaopeng are preparing to launch commercial aircraft in 2024 \cite{Ehang3248}.
These manufacturing, transportation, and related services involved in aircraft are becoming crucial economic growth points for regional industrial integration and employment.
A blue paper by Morgan Stanley estimates that the global UAM addressable market will reach US dollar $1.5$ trillion by 2040, which is on the same scale as the potential market of the autonomous vehicle \cite{Morgan2019}.

UAM is proposed to alleviate the ground passenger and cargo transportation loads in urban and suburban areas.
Authors in \cite{42358726Sunil} revealed that capacity was maximized when a layered airspace structure separated traffic at different flight altitudes.
Thus, UAM flight optimization should be devoted to optimizing safety, stability, and communication efficiency under the layered airspace.
However, there is a research blank in layered flight control regarding safe and efficient communication flight.

{
\subsection{RIS aided Communication}
RIS is a meta-surface that alters an incoming electromagnetic field in customized ways with low power consumption and hardware costs \cite{9140329Renzo}.
These custom-tailor features of RIS have attracted the interest from both academia and industry in studying the performance under various wireless systems such as massive multiple-input-multiple-output communication \cite{9875036Pap}, multi-access edge computing \cite{9693982Cao}, security communication \cite{10121733Naeem}, multi-hop RIS network \cite{9410457Huang}.
The authors in \cite{9110869Chongwen} joint optimized the transmission beamforming of the base station and the RIS through the DDPG method.
An Unmanned Aerial Vehicle (UAV) equipped with the RIS has been investigated in \cite{9416239Sixian} that functions as a passive relay to minimize the personal data rate received by the eavesdropper.
Cao \textit{et al.} \cite{9453804Cao} proposed a RIS-assisted transmission strategy to address the coverage and transmission rate problems for aerial-terrestrial communication systems.       
Although previous literature thoroughly researched the RIS signal construction capability, it remains in the communication performance to explore the RIS functions \cite{{9899454Xu},{Guo2_12416}}.
Previous work of RIS-UAV communication and flight control typically has no restrictions on the airspace structure\cite{{9749020Zhang}}. 
Therefore, the joint flight trajectory and communication optimization of RIS-aided aircraft dwells in an unresolved problem when it encounters structured airspace.
}

\subsection{Composite Potential Field}
The artificial potential field (APF) method utilizes virtual forces for trajectory control, which has local, straightforward, and real-time calculation advantages \cite{9143127Enming}.
Du \textit{et al.} \cite{10103224Guodong} proposed an enhanced APF method combined with an A-start algorithm to overcome the narrow path navigation problem. 
It verifies the field composition capability of different purpose forces. 
Xie \textit{et al.} \cite{9830995Songtao} investigated velocity and acceleration fields in that the composite APF algorithm can avoid overtaking in multi-vehicle systems. 
An improved APF is proposed in \cite{9538804Zhenhua} by introducing a rotating potential field to solve the problems of local minimum and oscillated solution. 
Fang \textit{et al.} \cite{7434002Fang} proposed a control scheme maintaining the communication connectivity of a UAV swarm, thus improving motion flexibility and reducing communication costs.
Jayaweera \textit{et al.} \cite{9234396Jay} proposed a novel dynamic artificial potential field path optimization technique to support UAVs-based ground-moving target tracking.
Although this APF scheme takes the communication connectivity into account, it ignores the transmission metrics in the composite APF design.
Very few studies on APF involve transmission rate into the scope of potential field design, and all assume that communication resources are sufficient, which is too idealistic.
\\

Overall, there is a research gap in layered trajectory optimization considering communication metrics. Moreover, RIS envisioned as a potential communication technology for UAM communication, still needs to explore its contributions to aircraft flight trajectory optimization in UAM with specific airspace.

\section{RIS-aided UAM Communication}

\begin{figure}[h]
\centering
     \includegraphics[width=.45\textwidth]{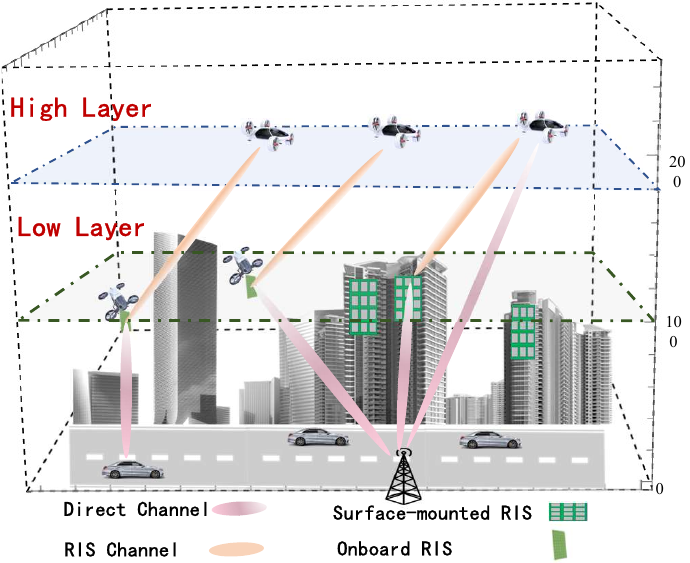} 
     \caption{Demonstration of the layered UAM.} 
\label{strucutre}
\end{figure}

Sunil \textit{et al.} \cite{42358726Sunil} verified that the vertical layered airspace benefits air traffic safety and capacity.
Therefore, Fig.~\ref{strucutre} exhibits a schematic of the layered UAM structure where
the height difference between adjacent vertical layers is $100 m$.
To ensure traffic safety within the layered structure, all aircraft undergoing the same layer must fly with the same velocity.
The prescribed velocity raises with the layer altitude to ensure separation between adjacent layers.
As shown in Fig.~\ref{strucutre}, short-travel aircraft can stay at low altitudes, i.e., the low layer. While long-travel aircraft can fast flight by flying at higher altitudes, i.e., the high layer.
These layered flight settings reduce the relative velocity of aircraft at the same layer, lessen the conflict, and improve overall traffic efficiency.

{
Typically, aircraft count on a predefined airway in traditional aviation systems. 
These airways are dictated by the wireless range of the ground-based ATCo \cite{2137469532745Lee}.
In addition, radio technologies have rapidly developed with the evolution of the 6G cellular network, which is considerably distinct from the previous.
Thus, it is high time for the industry and academia to re-design the UAM transportation system with novel 6G radio technologies.
\begin{figure*}
\centering
     \includegraphics[width=.95\textwidth]{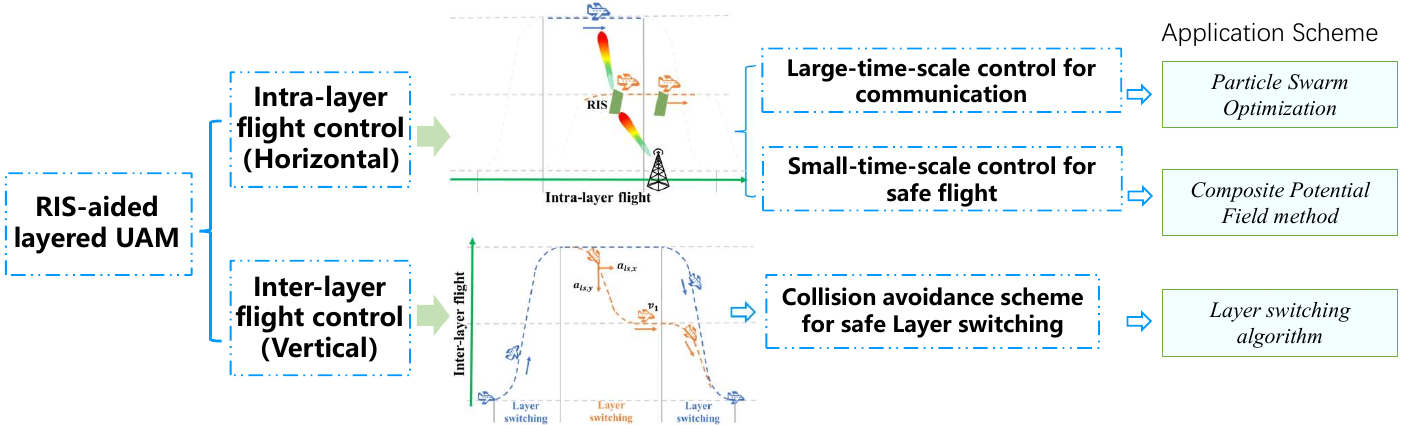} 
     \caption{Framework of RIS-aided trajectory and communication optimization.} 
\label{framework_based}
\end{figure*}

This paper explores the emerging 6G metasurface technology, i.e., Reconfigurable Intelligent Surfaces (RIS), which manipulates the electromagnetic behavior of radio waves to customize the communication environment.
The total procedure of the RIS-aided trajectory optimization is illustrated in Fig.~\ref{framework_based}, which includes the intra- and inter-layer parts.
The intra-layer flight control represents the horizontal trajectory and communication optimization.
The inter-layer flight control is mainly focused on vertical trajectory optimization for safe layer-switching.

The intra-layer flight control involves air-ground RIS communication and horizontal flight safety as shown in Fig.~\ref{framework_based}. 
This is because the communication capacity of high-layer aircraft is determined by the position and RIS phase shift matrix of the low-layer aircraft, which both rely on horizontal flight control.
Horizontal flight control comprises the large-time-scale optimization for communication quality and the small-time-scale optimization for safe aviation.
The particle swarm optimization method is applied to optimize the positions of low-layer and high-layer aircraft for air-ground RIS communication in large-scale optimization. 
Based on the optimal aircraft positions for RIS communication, we leverage the composite potential field method to optimize flight safety and communication in the small-time-scale optimization, simultaneously.

While the inter-layer flight performs the layer-switching procedure. The main concern of the layer-switching procedure is flight safety rather than communication performance. This is because Sunil \textit{et al.} \cite{42358726Sunil} pointed out that the duration of layer switching behavior should be minimized and the switching frequency also needs to be reduced as much as possible for safe aviation. 
This means that the communication optimization during this temporary and low-occurrence switching process is not noteworthy. However, in inter-layer flight control, layer-switching safety is more worthy of attention, as layer-switching is accompanied by considerable velocity changes that can easily lead to collisions. Thus, inter-layer optimization only involves the layer-switching flight algorithm to ensure flight safety in the vertical dimension.
}

\subsection{Dual-plane RIS Communication}
Specifically, we propose a dual-plane RIS communication scheme for layered UAM, in which all aircraft are equipped with a dual-channel communication module. 
This module engages low-frequency omnidirectional broadcasting with all participants on the control plane. 
Also, it utilizes high-frequency directional signals (such as mmWave, Terahertz, etc.) for target RIS communication on the data plane. 
\begin{figure}[h]
\centering
     \includegraphics[width=.4\textwidth]{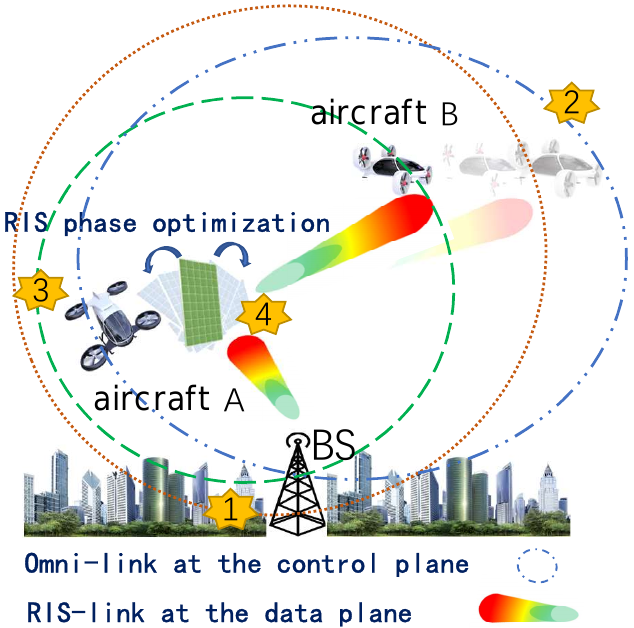} 
     \caption{Dual-plane RIS communication.} 
\label{dual_channel_scheme}
\end{figure}

The demonstration of the dual-plane communication is shown in Fig.~\ref{dual_channel_scheme}, where the control information (Request-To-Send, Clear-To-Send, RIS phase shift control, etc.) and kinetic information (aircraft velocity, position, etc.) are shared by the omnidirectional broadcasting at the control plane. 
Due to the attributes of low-frequency omnidirectional signals, the transmission range of the control plane is extensive. 
This paper regards that all participant aircraft and base stations can access the control plane.
But, the transmission rate of the omnidirectional signals is limited.
The control plane can only transmit a small volume of formatted data, like the control and kinetic information of the base station and aircraft.

Conversely, the base station employs high-frequency directional signals (mmWave and Terahertz) to transmit volume streams at the data plane.
Since the high directional/frequency signal has a high attenuation attribute during propagation, the transmission range of the data plane is confined.
As a consequence, the control and data planes can complement and benefit from each other.
Specifically, lite messages are transmitted through the control plane, while bulk messages pass through the data plane.

\begin{figure}[h]
\centering
     \includegraphics[width=.4\textwidth]{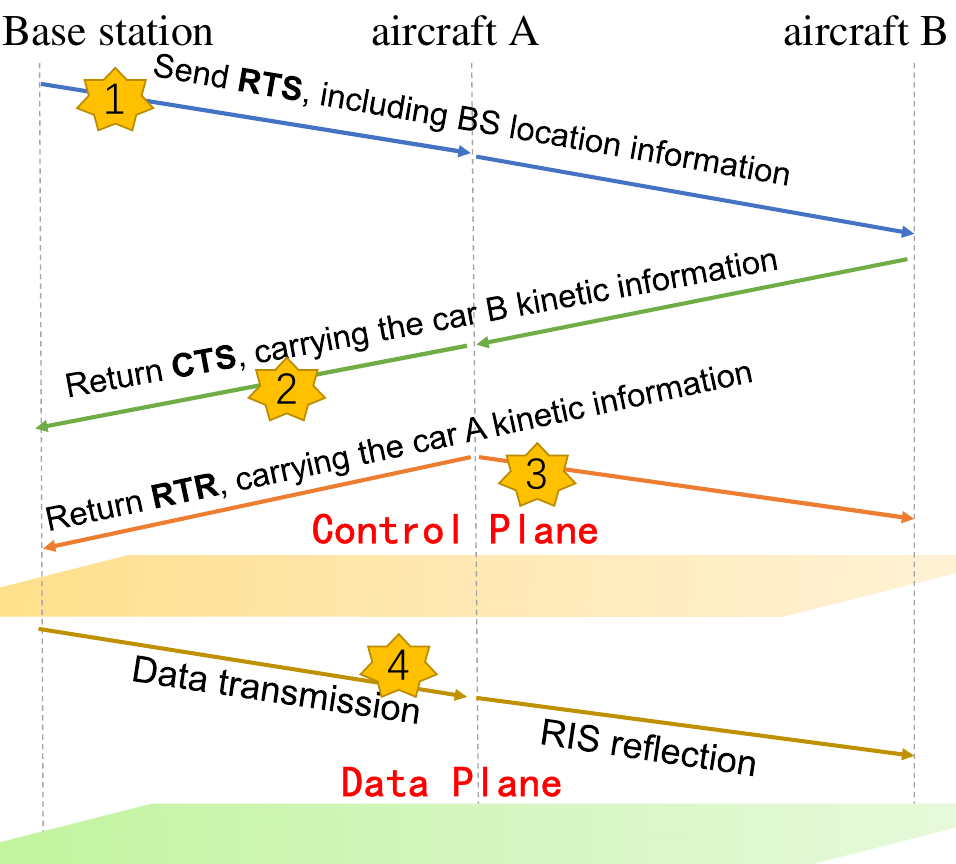} 
     \caption{Sequence diagram of the Dual-plane RIS communication.} 
\label{Seq_diagram_dualRIS}
\end{figure}

Additionally, Fig.~\ref{Seq_diagram_dualRIS} illustrates the diagram of the dual-plane RIS communication, in which the base station initially sends navigation and entertainment messages to aircraft B in the high layer.
Due to the signal attenuation, aircraft B cannot create a direct high-rate link with the base station.
The base station first delivers the Request-To-Send (RTS) to aircraft B in the omnidirectional channel at the control plane.
As aircraft B receives the RTS, it returns a Clear-To-Send (CTS) message to the base station through the control plane.
The RTS and CTS contain the status and kinetic information of the base station and aircraft B, respectively.

Since the base station and aircraft B both broadcast messages on the control channel, the nearby aircraft A will receive RTS and CTS messages at the control plane. 
In this scenario, aircraft A with the RIS panel aviates in the low layer.  
Consequently, aircraft A can calculate the RIS optimal phase shift based on the kinetic information of the base station and aircraft B attached in the RTS and CTS.
Once aircraft A produces the optimal RIS phase shift, it will send the Ready-To-Relay (RTR) message to notify the base station and aircraft B at the control plane.

The above protocol negotiations operate on the control plane, conveying the aircraft's kinetic and RIS status information.
When the negotiation has succeeded, the base station sends service data to aircraft B through the RIS of aircraft A. 
The serial labels in Fig.~\ref{Seq_diagram_dualRIS} correspond to those in Fig.~\ref{dual_channel_scheme}, which expresses the recalling flow order in the RIS communication.
Next, we will analyze the upper bound delay of RIS communication by network calculus.

\subsection{Performance Analysis of Dual-plane RIS Communication}
This subsection gives the upper bound delay $D(t,t+\Delta t)$ of the dual-plane RIS communication according to the sequential procedure as Fig.~\ref{Seq_diagram_dualRIS}.
Based on the network calculus theory, we get the close form of the control plane delay, data plane delay, and dual-plane RIS communication delay, respectively.

In network calculus theory, an arrival process $A(\tau,t) = A(t) - A(\tau)$ represents the cumulative volume of the network traffic from the base station to aircraft B during $(0,t]$ interval \cite{6868978Fidler}.
The convolution $\otimes$ of network calculus is the min-plus convolutions, which is defined as,
\begin{equation}
\begin{split}
\begin{aligned}
(a \otimes b)(x) = \inf_{0 \leq y \leq x} \left[ a(y) + b(x - y). \right].
\end{aligned}
\end{split}
\label{sfdigulsdhfgl}
\end{equation}

We assume the network traffic generation follows the Poisson process.
Therefore, $A(t) = \frac{(\lambda t)^n}{n!}e^{-\lambda t}$.
The service curve of the wireless channel is modeled as a classic latency rate curve $\beta(t)=r\times (t-T)^+$ where $r$ is the transmission rate, and $T$ is the access and message queue delay. $(x)^+=\max\{0,x\}$ \cite{Jiang2008Stochastic}.

First of all, we investigate the transmission procedure at the control plane. 
Since all participants are connected to the control plane, they can directly communicate with each other in one hop by the omnidirectional signal.
The channel service curve of the control plane $\beta_{omni}(t)$ is,
\begin{equation}
\begin{split}
\begin{aligned}
\beta_{omni}(t) = r_{omni} \times (t- T_{omni})^+,
\end{aligned}
\end{split}
\label{_w_ripuewaifgw}
\end{equation}
\noindent where $r_{omni}$ indicates the transmission rate of the omnidirectional channel. 
$T_{omni}$ is the access and message queue delay. 
To simplify the analysis, we approximate the $T_{omni}=\zeta \frac{l_{data}}{r_{omni}}$ where $l_{data}$ represents the volume of the data message and $\zeta$ is a weight parameter.

Regarding the data plane, due to the traits of the high-frequency directional signal, direct transmission without RIS at the data plane needs additional RTS- and CTS-confirmed procedures for accurate alignment.
Thus, the stages of the data plane transmission comprise three parts: the RTS sending, the CTS response, and the data transmission. 
Then, the corresponding channel service curve $\beta_{direct}(t)$ of the control plane transmission is,
\begin{equation}
\begin{split}
\begin{aligned}
\beta_{direct}(t) &= \beta_{RTS}(t) \otimes \beta_{CTS}(t) \otimes \beta_{d}(t) \\
&= \min\{r_{RTS}, r_{CTS}, r_{d}\} \times (t-T_{d})^+,
\end{aligned}
\end{split}
\label{sdfjhweurf2qwetyu}
\end{equation} 

\noindent in which $\beta_{RTS}$, $\beta_{CTS}$, and $\beta_{d}$ refer to the service curve of RTS sending, CTS response, and data transmission using the direct high-rate channel, respectively. 
Based on the min-plus convolution law, the access delay $T_{d}$ of data plane is calculated as $T_{d} =\zeta [ \frac{l_{RTS}}{r_{RTS}} +  \frac{l_{CTS}}{r_{CTS}} +  \frac{l_{data}}{r_{d}}  ]$, where $l_{RTS}$ and $l_{CTS}$ are the volume of RTS and CTS message, respectively \cite{6868978Fidler}.
In general, RTS and CTS messages are both transmitted at the control plane by the omnidirectional signal.
It is appropriate to set that the transmission rate of RTS $r_{RTS}$ and CTS $r_{CTS}$ equal to the omnidirectional transmission rate, i.e., $r_{RTS} = r_{CTS} = r_{omni}$.

Furthermore, the dual-plane RIS communication follows the stage procedure of Fig.~\ref{Seq_diagram_dualRIS}.
According to the cascading rule of network calculus, this RIS transmission procedure is expressed as \cite{6868978Fidler},
\begin{equation}
\begin{split}
\begin{aligned}
&\beta_{RIS}(t) \\&= \beta_{RTS}(t) \otimes \beta_{CTS}(t) \otimes \beta_{RTR}(t) \otimes \beta_{\eta_1}(t) \otimes \beta_{\eta_2}(t)\\
&= \min\{r_{RTS}, r_{CTS}, r_{RTR}, r_{\eta_1}, r_{\eta_2}\} \times (t-T_{RIS})^+,
\end{aligned}
\end{split}
\label{vuyresgdfsuuysdriv}
\end{equation} 

\noindent where $\beta_{RTR}$, $\beta_{\eta_1}$, $\beta_{\eta_2}$ are the service curve of RTR broadcasting channel, base station-to-RIS channel, and RIS-to-aircraft channel, respectively.
Also, the RTR transmission operates at the control plane, i.e., the transmission rate $r_{RTR}$ of the RTR message is identical to $r_{omni}$. 
Thus, $T_{RIS} =\zeta [ \frac{l_{RTS}}{r_{RTS}} +  \frac{l_{CTS}}{r_{CTS}} +  \frac{l_{RTR}}{r_{RTR}} +  \frac{l_{data}}{r_{\eta_1}} +\frac{l_{data}}{r_{\eta_2}} ]$.
Hereafter, according to the network calculus, the upper bound delay is defined as,
\begin{equation}
\begin{split}
\begin{aligned}
D(t) \leq \min \left\{\omega \geq 0: \max _{\tau \in[0, t]}\left\{A(\tau, t)-\beta(\tau, t+\omega)\right\} \leq 0\right\},
\end{aligned}
\end{split}
\label{vuyresgdfsuuys12iv}
\end{equation} 

{
Suppose that a network traffic with a fixed unit packet size follows a Poisson process with mean rate $\lambda$. 
Based on the stochastic network calculus, we can get the statistical delay upper bound of the omnidirectional channel,
\begin{equation}
\begin{split}
\begin{aligned}
&P\{D(t)>t\}=\\&P\{A(\tau, \tau+t)-\lambda t> \beta \} \\
&\leq
\sum_{k=\lceil \beta +\lambda t\rceil}^{\infty}\left\{\frac{e^{-\lambda t}[\lambda t]^k}{k !}\right\},
\end{aligned}
\end{split}
\label{vuyresgdfsuuysdriv}
\end{equation} 
 
\noindent where $D(t)$ is the delay upper bound. 
$P\{D(t)>t\}$ represents that the transmission delay $t$ is less than the upper bound $D(t)$. We name $P\{D(t)>t\}$ as the delay valid probability.
Therefore, the transmission failure probability, i.e., exceeded the delay upper bound $D(t)$, is $1-P\{D(t)>t\}$.

In dual-plane RIS communication, control packets such as RTS, CTS, and RTR may experience packet loss, resulting in transmission failure or a large data transmission delay.
Assuming the probability of packet loss is $(1-p)$ and the probability of successful transmission is $p$. 
Suppose BS or aircraft does not receive the corresponding acknowledgment (ACK) packet after waiting for a specific Time-To-Live (TTL) threshold. In that case, it indicates that the control packet has been lost and the lost packet re-transmission is triggered immediately.

For the direct high-rate channel of the data plane, the delay of direct high-rate channel $D_{direct}$ is composed of $D_{direct}=D_1 + D_2 + D_s^{dir}$ where $D_1$ is the delay caused by the RTS packet loss, $D_2$ is the delay caused by the CTS packet loss, and $D_s^{dir}$ is the delay of successful control message transmission on the direct channel.
Assuming the control packet is successfully sent after the $k$-th retransmission. We get $D_1=(k-1) t_{RTS}$ and $D_2=(k-1) t_{CTS}$, where $t_{RTS}$ represents the TTL threshold for RTS and $t_{CTS}$ is the TTL threshold for CTS.

The probability of successful transmission at the $i$-th retransmission is ${P\{k=i\}=(1-p)^{i-1} p}$.
We derive the delay valid probability of $D_1$ as,
\begin{equation}
\begin{aligned}
&P\left\{D_1>t_1\right\} = P\left\{(k-1) t_{RTS}>t_1\right\} \\
&= P\left\{k>\frac{t_1}{t_{RTS}}+1\right\} = 1-P\left\{k \leq \frac{t_1}{t_{RTS}}+1\right\} \\
&= 1-\sum_{i=1}^{\left\lceil\frac{t_1}{t_{RTS}}+1\right\rceil}(1-p)^{i-1} p \\
&= (1-p)^{\left\lceil\frac{t_1}{t_{RTS}}+1\right\rceil }\triangleq f_{D_1}.
\end{aligned}
\label{vuyresgdfsuuysdriv}
\end{equation}

\noindent Similarly, the delay valid probability of $D_2$ is,
\begin{equation}
\begin{aligned}
&P\left\{D_2>t_2\right\} = P\left\{(k-1) t_{CTS}>t_2\right\} \\
&= P\left\{k>\frac{t_2}{t_{CTS}}+1\right\} \\
&= (1-p)^{\left\lceil\frac{t_2}{t_{CTS}}+1\right\rceil}\triangleq f_{D_2}.
\end{aligned}
\label{vuyresgdfsuuysdriv}
\end{equation}

\noindent In addition, the delay $D_s^{dir}$ valid probability of the direct channel is,
\begin{equation}
\begin{split}
\begin{aligned}
P\{D_s^{dir} > t_s \}\leq
\sum_{k=\lceil \beta_{d} +\lambda t_s\rceil}^{\infty}\left\{\frac{e^{-\lambda t_s}[\lambda t_s]^k}{k!}\right\}\triangleq f_{D_s^{dir}}.
\end{aligned}
\end{split}
\label{vuyresgdfsuuysdriv}
\end{equation} 
\noindent where $\beta_{d}$ is the service curve of the direct channel.
Moreover, according to \cite{Jiang2008Stochastic}, there is,
\begin{equation}
\begin{split}
\begin{aligned}
\bar{F}_Z(x) \leq \bar{F}_{X_1} \otimes \bar{F}_{X_2} \otimes \ldots \otimes \bar{F}_{X_N}(x),
\end{aligned}
\end{split}
\label{vuyresgdfsuuysdriv}
\end{equation} 

\noindent in which $Z$ and $X_i$ are random variables, and $Z=\sum_{i=1}^N X_i$. $\bar{F}_Z(x)$ represents the complementary cumulative distribution function (CCDF) of $Z$ with respect to variable $x$. 

{
Consequently, the upper bound of the delay valid probability of the direct channel is given as,
\begin{equation}
\begin{aligned}
& P\{D_{direct}>t\} = P\left\{D_1+D_2+D_s^{dir}>t\right\} \\
&\leq f_{D_1} \otimes f_{D_2} \otimes f_{D_s^{dir}}\\
&= \inf_{t_1+t_2+t_s=t} \left\{(1-p)^{\left\lvert \frac{t_1}{t_{RTS}}+1 \right\rvert} 
    + (1-p)^{\left\lvert \frac{t_2}{t_{CTS}}+1 \right\rvert}\right. \\
&\quad \left. + \sum_{k=\left[\beta_{\text{direct}}+\lambda t_s\right]}^{\infty} \left\{\frac{e^{-\lambda t_s}\left(\lambda t_s\right)^k}{k!}\right\} \right\}
\end{aligned}
\label{vuyresgdfsuuysdriv}
\end{equation}

For RIS channel, the delay upper bound $D_{RIS}$ is $D_{RIS}=D_1 + D_2 +D_3+D_s^{RIS}$, where $D_1$ is the delay caused by the RTS packet loss, $D_2$ is the delay caused by the RTS packet loss, $D_3$ is the delay caused by the RTR packet loss, and $D_s^{RIS}$ is the delay of successful control message transmission on the RIS channel.
Assuming the control packet is successfully sent after the $k$-th retransmission. 
There is $D_3=(k-1) t_{RTR}$ where $t_{RTS}$ represents the TTL threshold for RTR.

Therefore, Similar to the previous analysis, the delay valid probability caused by RTR packet loss is,
\begin{equation}
\begin{aligned}
&P\left\{D_3>t_3\right\} = P\left\{(k-1) t_{RTR}>t_3\right\}\\
&= (1-p)^{\left\lceil\frac{t_3}{t_{RTR}}+1\right\rceil}\triangleq f_{D_3}. 
\end{aligned}
\label{vuyresgdfsuuysdriv}
\end{equation}

\noindent The delay $D_s^{RIS}$ valid probability of the RIS channel is,
\begin{equation}
\begin{split}
\begin{aligned}
P\{D_s^{RIS}>t_s \}&\leq
\sum_{k=\lceil \beta_{RIS} +\lambda t_s\rceil}^{\infty}\left\{\frac{e^{-\lambda t_s}[\lambda t_s]^k}{k !}\right\} \\
&\triangleq f_{D_s^{RIS}},
\end{aligned}
\end{split}
\label{vuyresgdfsuuysdriv}
\end{equation} 
\noindent where $\beta_{RIS}$ is the service curve of the RIS channel.
As a consequence, the upper bound of the delay valid probability of the RIS channel is,
\begin{equation}
\begin{aligned}
& P\{D_{RIS}>t\} = P\left\{D_1+D_2+D_3+D_s^{RIS}>t\right\} \\
&\leq f_{D_1} \otimes f_{D_2} \otimes f_{D_3}\otimes f_{D_s^{RIS}}\\
&= \inf_{t_1+t_2+t_3+t_s=t} \left\{(1-p)^{\left\lceil \frac{t_1}{t_{RTS}}+1 \right\rceil} 
    + (1-p)^{\left\lceil \frac{t_2}{t_{CTS}}+1 \right\rceil} \right. \\
&\quad \left. + (1-p)^{\left\lceil \frac{t_3}{t_{RTR}}+1 \right\rceil} 
    + \sum_{k=\left[\beta_{\text{RIS}}+\lambda t_4\right]}^{\infty} \left\{\frac{e^{-\lambda t_4}\left(\lambda t_4\right)^k}{k!}\right\} \right\}
\end{aligned}
\label{vuyresgdfsuuysdriv}
\end{equation}

In the Simulation section, we verify the delay upper bound for the different transmission fashions, among which the dual-plane RIS communication has the lowest delay upper bound and transmission failure probability.
}
}
\section{RIS-aided Trajectory Optimization}
This section presents the joint optimization of communication and trajectory for layered aviation.
Specifically, we first depict the aircraft behavior in the layered UAM.
There are three layers in the UAM, namely ground, low and high layers. $N={0, 1, 2}$ is the layer indicator from ground $N = 0$ to the high layer $N=2$ in order.
According to the analysis of Section III, all aircraft within the same layer are required to keep the same velocity. 
And the velocity increases with the layer indicator $N$, i.e., the altitude.

The passenger travel demand and communication quality determine the flight trajectory of an aircraft.
The trajectory design contains layer selection, layer switching strategy, and velocity control.
We assume all aircraft are mounted with the RIS panel on their surface.
Therefore, the lower-layer aircraft trajectory can affect the upper-layer aircraft communication quality.
This scenario entangles trajectory optimization with communication performance, increasing the difficulty of joint communication and trajectory optimization.
To address the challenge, we propose the communication and trajectory joint optimization scheme for the layered UAM.

\subsection{Communication Optimization}
According to the dual-plane RIS communication procedure, the low-layer aircraft reflects the base station signal to the high-layer aircraft. 
When the transmission negotiation has been accomplished at the control plane, only one aircraft in the low layer is assigned to perform RIS reflection for the high-layer communication. 
Set the coordinates of the base station (BS) as $p_{BS} = [x_{BS}, 0]$. 
The distance between low-layer aircraft $i$ and BS is written as $d_{BS, i}$.
The direct connection channel from the base station to the high-layer aircraft $k$ is,
\begin{equation}
\begin{split}
\begin{aligned}
h_{BS, k}=\sqrt{\frac{\beta}{\left(d_{B S, k}\right)^{\alpha_{B S, k}}}} \sqrt{\frac{\kappa^d}{\kappa^d+1}} h_{BS, k}^{LoS},
\end{aligned}
\end{split}
\label{DFSJKGHSBREUYGBDF}
\end{equation}
\noindent in which $\beta$ is the reference channel coefficient, $\alpha_{B S, k}$ represents the path loss index.
Denote by $\kappa^d$ the Rician factor. 
Thus, $h_{BS, k}=\sqrt{\frac{\beta}{\left(d_{B S, k}\right)^{\alpha_{B S, k}}}}  h_{BS, k}^{LoS}$. 
$ h_{BS, k}^{LoS}=1$ indicates a deterministic line-of-sight (LoS) component.
Additionally, the channel from the BS to the RIS mounted on low-layer aircraft $i$ is,
\begin{equation}
\begin{split}
\begin{aligned}
&h_{BS, i} =\sqrt{\frac{\beta}{\left(d_{B S, i}\right)^{\alpha_{BS, i}}}} \\
&\times   \left[1,   e^{-j \frac{2 \pi d_s}{\lambda} \cos \varphi_{BS, i}}  \cdots,   e^{-j \frac{2 \pi(\sqrt{L}-1) d_s}{\lambda} \cos \varphi_{B S, i}}\right],
\end{aligned}
\end{split}
\label{dfsbikujsdfkb}
\end{equation}
\noindent where $\lambda$ is the wavelength, $d_s$ is the spacing between adjacent meta-components, and $L$ is the number of meta-components arranged in a square RIS panel.
Denote by $\varphi_{BS, i}$ the signal incident angle, i.e., $\cos \varphi_{BS, i} = \frac{x_i - x_{BS}}{d_{BS,i}}$, wherein $x_i$ and $x_{BS}$ are the horizontal coordinates of low-layer aircraft $i$ and BS, respectively.
Similarly, the channel from the RIS of the low-layer aircraft $i$ to the high-layer aircraft $k$ is given as,
\begin{equation}
\begin{split}
\begin{aligned}
h_{i, k}=\sqrt{\frac{\beta}{\left(d_{i, k}\right)^{\alpha_{i, k}}}} h_{i, k}^{LoS}.
\end{aligned}
\end{split}
\label{dfguyertsuy}
\end{equation}

\noindent $ h_{i, k}^{LoS}$ is a line-of-sight (LoS) component of RIS-to-aircraft $k$ channel. Therefore,
\begin{equation}
\begin{split}
\begin{aligned}
&h_{i,k}^{LoS} =\left[1,   e^{-j \frac{2 \pi d_s}{\lambda} \cos \varphi_{i,k}}  \cdots,   e^{-j \frac{2 \pi(\sqrt{L}-1) d_s}{\lambda} \cos \varphi_{i,k}}\right],
\end{aligned}
\end{split}
\label{fgdhnuyreyy}
\end{equation}
\noindent The signal-to-noise ratio is given as,
\begin{equation}
\begin{split}
\begin{aligned}
r_{BS, i, k}=\frac{\left|h_{BS, k}^H+h_{BS, i}^H \Theta_i h_{i, k}\right|^2 P_{BS}}{\sigma^2},
\end{aligned}
\end{split}
\label{dcxvdsfuighrtsi}
\end{equation}

\noindent in which, $P_{BS}$ is the transmission power. $\Theta_i$ is the RIS phase shift matrix of aircraft $i$, i.e., 
\begin{equation}
\begin{split}
\begin{aligned}
\Theta_i=diag (e^{j\theta_{i,1}}, e^{j\theta_{i,2}}, \cdots, e^{j\theta_{i,L}}).
\end{aligned}
\end{split}
\label{jkjktuut}
\end{equation}
And $\theta_{i,l}\in [0, 2 \pi], l =\{1,2,\dots,L\}$. To maximize the communication capacity of BS to high-layer aircraft $k$, we give the communication optimization problem as,
\begin{equation}
\begin{aligned}
\begin{split}
 \textbf{P1:} \qquad &\max \sum_{t=\Delta t}^\mathcal{T} \sum_{i=1}^M   C_{BS,i}(t)\\
 \text{s.t.} \ \ \ \ \ 
&\text{C1:} \ \ \theta_{i,l}\in [0,2\pi], 
\end{split} 
\end{aligned} 
\label{sdffff}
\end{equation}
\noindent where $C_{BS,i,k}=B\log_2 (1+r_{BS,i,k} )$ is the communication capacity. 
$M$ is the total number of aircraft. 
$\mathcal{T}$ represents the investigated flight duration.
$\Delta t$ is the minimum time interval for communication optimization, larger than the upper bound delay $D(t)$ of RIS communication in section III.
Substituting $d_s= \frac{\lambda}{2}$ into $P1$, the optimal solution satisfies that,
\begin{equation}
\begin{aligned}
\begin{split}
&h_{B S, i}^H \Theta_i h_{i, k}=\\
&\frac{\beta   \left(e^{j \theta_{i, 1}}+\cdots+e^{j\left(\pi(\sqrt{L}-1)\left(\cos \varphi_{B S, i}-\cos \varphi_{i, k}\right)+\theta_{i, L}\right)}\right)}{\sqrt{\left(d_{B S, i}\right)^{\alpha_{B S, i}}\left(d_{i, k}\right)^{\alpha_{i, k}}}},
\end{split} 
\end{aligned} 
\label{sdajfyhgasfkgeiwruygfwu}
\end{equation}

\noindent where $\varphi$ is the cosine of elevation angle that is,
\begin{equation}
\begin{aligned}
\begin{split}
\cos \varphi_{BS,i}=\frac{x_i-x_{BS}}{d_{BS,i}},
\end{split} 
\end{aligned} 
\label{vyryryruwk}
\end{equation}

\noindent in which, ${d_{BS,i}}  = ||  s_i - s_{BS} ||$. $s_i$ and $s_{BS}$ are the position of the aircraft $i$ and base station, respectively. 
$x_i$ and $x_{BS}$ are the x-coordinate of $s_i$ and $s_{BS}$, respectively.
Therefore, we obtain the optimal RIS phase shift matrix $\theta^*_{i,l}, \  l\in\{1,\cdots, L\}$ maximizing $P1$ as,
\begin{equation}
\begin{aligned}
\begin{split}
\theta^*_{i,l}=-\pi\left[(l-1)\bmod{\sqrt{L}}\right](\cos \varphi_{BS, i} - \cos\varphi_{i,k}).
\end{split} 
\end{aligned} 
\label{dfgsuhryueueryry}
\end{equation}

Additionally, due to the fabrication constraints of the magnetic micro-actuator, $\theta_{i,l}$ is typically a discrete value. 
This discrete RIS phase shift control will impact the trajectory optimization of aircraft.
Consequently, the optimal communication positions are sequence discrete spots. 
In the next section, we will engage these sequence discrete spots in the intra-layer trajectory optimization.

\subsection{Trajectory Optimization}
Flight trajectory optimization takes flight safety, stability, and communication efficiency into account simultaneously.
At first, we explore the safety separation in the layered airspace to provide flight safety.
In layered UAM, the safety separation between adjacent aircraft comprises horizontal and vertical separation.

\begin{figure}[h]
\centering
     \includegraphics[width=.48\textwidth]{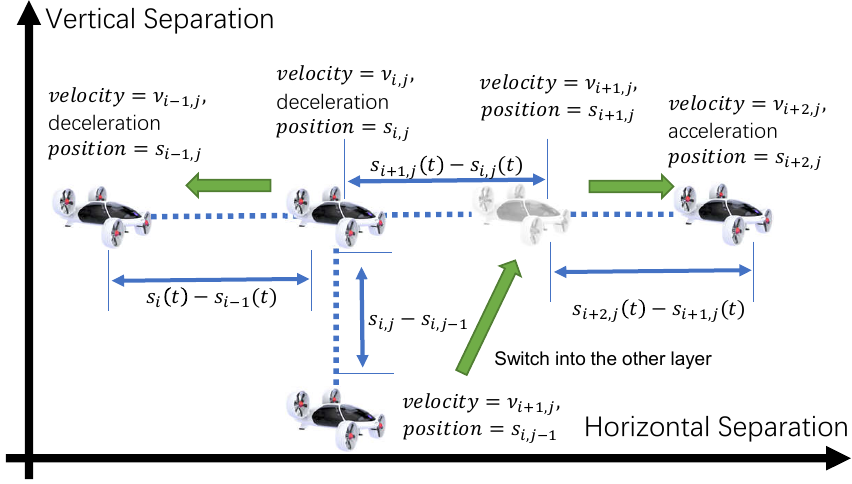} 
     \caption{Two-dimensional safety separations in layered UAM.} 
\label{safety_separation}
\end{figure}

As demonstrated in Fig.~\ref{safety_separation}, if the distance of adjacent aircraft in the same layer is less than the horizontal safety separation, the layer-switching behavior of the aircraft will be triggered to avoid horizontal collision accidents. 
Specify the position of the aircraft $i$ at time $t$ as $s_i(t)=[x_i(t), h_i(t)]$, with a velocity of $v_i(t)=[v_{i, x}(t), v_{i, y}(t)]$ where $v_{i, x}(t)$ represents the horizontal velocity and $v_{i, y}(t)$ is the vertical velocity.
The horizontal safety separation $d^H_{safe}$ is a proper spacing for collision avoidance of adjacent aircraft in the same layer.

In the vehicle following scenario, the preceding aircraft starts to brake with the deceleration $B$ at time $t=0$. The follower detects the braking behavior at $t=t_{1}$. 
Then, it slows down with the deceleration $b$ at $\tau=t_{2} + t_{1}$.
Wherein $t_{1}$ is the interval from the braking of the preceding aircraft to the follower perceived.
$t_{2}$ is the time consumed by the on-board processing and reaction of the follower. 
To avoid accidents, the horizontal safety separation $d^H_{safe}$ between adjacent aircraft caters to Eq.~(\ref{fdgbshbuiorteh}) at $t=0$,
\begin{equation}
\begin{aligned}
\begin{split}
d^H_{safe}=\frac{B-b}{2Bb} v_{n_i}^2+v_{n_i} (t_1+t_2).
\end{split} 
\end{aligned} 
\label{fdgbshbuiorteh}
\end{equation}

\noindent Note that $\tau = t_{2}+t_{1}$ refers to the perception-reaction delay, which is the duration of time from an emergency happens to the aircraft reaction \cite{Nekoui2010Fundamental}. 
Horizontal flight safety is guaranteed as the distance between adjacent aircraft is larger than the safety separation.

While two aircraft are at different layers, the vertical safe separation is given as,
\begin{equation}
\begin{aligned}
\begin{split}
d^V_{safe}=c^{vert}_{safe} ||v_{n_i}|| \cos\gamma_{i,j},
\end{split} 
\end{aligned} 
\label{fwrdgbfgfuoysdf}
\end{equation}
\noindent where $c^{vert}_{safe}$ is a constant coefficient. $v_{n_i}$ is the velocity of the aircraft $i$ in the layer $n$. $\gamma_{i,j}$ is the angle between the position vector and velocity vector, i.e.,
\begin{equation}
\begin{aligned}
\begin{split}
\cos\gamma_{i,j}=\frac{-s_{i,j} \cdot v_{i,j}}{||s_{i,j}|| ||v_{i,j}||},
\end{split} 
\end{aligned} 
\label{fwrdgbfgfuoysdf}
\end{equation}

\noindent in which $s_{i,j}=s_i-s_j$ and $v_{i,j}=v_i-v_j$. Hence, the safety separation and flight behavior are different in the horizontal and vertical directions.
Besides, the kinetic parameters of aircraft in the horizontal and vertical directions are orthogonal and do not involve each other.
Consequently, we can separately design the intra-layer (horizontal) and inter-layer (vertical) trajectory optimization.
Wherein, the intra-layer trajectory optimization primarily deals with the flight courses of aircraft within the same layer. 
In contrast, inter-layer trajectory optimization focuses on the layer-switching operation of aircraft.
Hereafter, we will detail the solutions of these two optimizations.

\subsubsection{ \textbf{Intra-layer Trajectory Optimization}}
We assume the total number of aircraft in the layered UAM is $M$ during a specific period $\mathcal{T}$.
In addition, an aircraft in a layer only adopts the one-way straight-flight driving behavior.
The reason is that the turning behavior may threaten the flight safety of aircraft with high velocity in the horizontal direction. 
Further exploration of safe turning regulations is out of the scope of this paper.
Therefore, the horizontal trajectory optimization problem for flight safety and communication efficiency is formed as,
\begin{equation}
\begin{aligned}
\begin{split}
 \textbf{P2:} \qquad &\min \sum_{i=1}^M \sum_{t=\Delta t}^\mathcal{T}  w_1(d_{i,pre}-d_{safe})^2\\
 &+w_2\left(  (v_{i,x} - v_{l})^2 + v_{i,y}^2 \right)  - w_3  C_{BS,i}\\
 & + || s_i(t) - s_{goal} ||\\
 \text{s.t.} \ \ \ \ \ 
 &\text{C1:} \ \ \theta_{i,l}\in [0,2\pi], \\
&\text{C2:} \ \ d_{i,j}\ge d_{safe}, \\
&\text{C3:} \ \ h_{n,i} =  n H, \\
&\text{C4:} \ \ v_{2} > v_{1} > v_{0},\\
&\text{C5:} \ \ || s(t+\Delta t) - s(t) || \le v_{max} \Delta t,
\end{split} 
\end{aligned} 
\label{fsdgb8yousederao8uogfhae}
\end{equation}
\noindent wherein the position $d$, velocity $v$, and communication capacity $C$ are both functions with time $t$. 
To clarify the derivation, we omit the time variable in these expressions.
The optimization variables include each aircraft position $s$, acceleration $a_i(t)$, and the RIS phase shift $\theta_{i, k}(t)$. 
Moreover, $v_{l}$ is the expected velocity of aircraft in layer $l$.
$v_0$ represents the expected velocity on the ground.
$v_1$ and $v_2$ are the expected velocities in the low layer and high layer, respectively.
According to \cite{42358726Sunil}, these expected velocities are constants specified by traffic managers based on traffic conditions. 
Constraint $C3$ describes the layered structure constraint, where $h_{n,i}$ is the height of aircraft $i$ in the $n$-th layer.
Constraint $C4$ represents that the expected velocity of the high-layer aircraft is larger than that of the low-layer aircraft.
And the velocity of the aircraft is higher than its ground speed.

To simplify the solution of $P2$, we can first investigate the optimal solution of a single aircraft in a time slot $\Delta t$. 
When the trajectory of an aircraft in each time slot is optimal, overall optimization is also achieved.
Thus, the joint optimization problem can be converted to,

\begin{equation}
\begin{aligned}
\begin{split}
 \textbf{P3:} \qquad &\min w_1(d_{i,pre}-d_{safe})^2\\
 &+w_2\left(  (v_{i,x} - v_{n_i})^2 + v_{i,y}^2 \right)  - w_3  C_{BS,i}\\
 \text{s.t.} \ \ \ \ \ 
 &\text{C1}, \ \text{C2}, \ \text{C3}, \ \text{C4}.
\end{split} 
\end{aligned} 
\label{dsfvhbgyurstgytreye}
\end{equation}

\noindent To address this problem, an iterative dual-time-scale optimization scheme is proposed to find a sub-optimal solution by employing the Particle Swarm and Composite Potential Field (CPF) methods.
Moreover, the optimization variables are decomposed into several blocks.
Each block optimizes a single-variable problem as other parameters are settled. 
Specifically, $P3$ is divided into the communication-oriented trajectory optimization part and the safety-oriented trajectory optimization part.

We propose a dual-time-scale trajectory method to solve the block problems iteratively. 
The communication-oriented trajectory optimization is derived on a large time scale. 
In contrast, the safety-oriented trajectory optimization is applied on a small time scale.
This is because flight safety is more important than that of communication quality. 
Hence, safe trajectory optimization requires more frequent updates than communication. 
It implies that we need to fulfill the safe trajectory optimization in a small time slot.

  \emph{1a) Large-time-scale Trajectory Optimization:}

The optimal RIS phase shift matrix $\Theta$ is given by Eq.~(\ref{dfgsuhryueueryry}).
However, the practical RIS phase shift can only take discrete values due to manufacturing limitations.
Thus, we suppose that the phase shift $\Theta$ only takes the discrete value $\xi\pi$ where $\xi$ is a rational number to determine the minimum resolution of discrete values. 
The calculated optimal $\theta$ should be close to the practical phase shift $m\xi\pi$ for actual implementation, where $m$ is an integer number. There is,

\begin{equation}
\begin{aligned}
\begin{split}
&\min \  \left(m\xi \pi- \theta \right)^2.
\end{split} 
\end{aligned} 
\label{yujuykyuk}
\end{equation}
\noindent Therefore, based on Eq.~(\ref{vyryryruwk}), (\ref{dfgsuhryueueryry}), and (\ref{yujuykyuk}), the optimal aircraft position for communication has the following objective,
\begin{equation}
\begin{aligned}
\begin{split}
&\min \  \left[ m\xi  -  \mathcal{U} \left(\frac{x_{1_i}^*(t)-x_{B S}}{\left\|s_{1_i}^*(t)-s_{B S}\right\|}-\frac{x_{2_k}^*(t)-x_{1_i}^*(t)}{\left\|s_{1_i}^*(t)-s_{2_k}^*(t)\right\|}\right)   \right]^2,
\end{split} 
\end{aligned} 
\label{dsfvhbgyurstgytreye}
\end{equation}

\noindent where $\mathcal{U} = (l-1)\bmod{\sqrt{L}}$. 
$s^*(t)$ is the optimal position to maximize the communication rate with a specific discrete phase shift $\theta$.
$x^*$ is the x-coordinate of $s^*$.
Thus, we can disassemble the $P3$ as the communication-oriented trajectory optimization problem,
\begin{equation}
\begin{aligned}
\begin{split}
 \textbf{P4:} \qquad &\min w_1^*\frac{1}{L} \sum\limits_{l=1}^L   \left( n\xi  -  \mathcal{U} \mathcal{J}   \right)^2 \\
\text{s.t.} \ \ \ \ \ 
 &\text{C1:} \ \ ||s_{n_i}(t+q\Delta t) - s_{n_i}(t)|| \le v_{max}q\Delta t, \\
&\text{C2:} \ \ h_{n,i}(t) = n H, \\ 
&\text{C3:} \ \ x_{s_{n_i}(t+q\Delta t) } - x_{s_{n_i}(t)} > 0, \\ 
\end{split} 
\end{aligned} 
\label{dfsbdfubyhsdfj}
\end{equation}

\noindent where $\mathcal{J} = \left(\frac{x_{1_i}^*(t)-x_{B S}}{\left\|s_{1_i}^*(t)-s_{B S}\right\|}-\frac{x_{2_k}^*(t)-x_{1_i}^*(t)}{\left\|s_{1_i}^*(t)-s_{2_k}^*(t)\right\|}\right)$.
It requires the aircraft to reach its final destination by passing several optimal communication positions.
And $x_{s_{n_i}(t)}$ is the $x$ coordinate of the position $s_{n_i}(t)$. 
$C1$ is the maximum velocity constraint.
$C2$ is the layered airspace constraint.
$C3$ indicates that aircraft always fly ahead.
Subsequently, we propose the particle swarm method to obtain the sub-optimal solution of the communication-oriented problem in a heuristic way.

The particle swarm algorithm is shown as Alg.~\ref{Ag1}.
First of all, we initialize the positions and velocities of all particles. 
In each iteration, calculate the fitness function of each particle, i.e., the optimization objective of $P4$. 
Then, update the fitness records of each particle as well as the particle velocity and position. 
The whole process stops while attaining the maximum iterations.

\begin{algorithm} 
    \caption{Particle Swarm Optimization} \label{Ag1}
    \textbf{Input}: positions of aircrafts $s_i(0)$, base station $s_{BS}$, and the number of RIS meta-elements $L$. 
    
    \textbf{Output}: optimal position $s_{i, goal}^*(t)$
    
    Initialize the positions and velocities of all particles;

    \For{$iter=1,2,\cdots,max\_iter$}{
        Calculate the fitness of each particle according to the objective of $P4$;\\

        Update the fitness of each particle;\\

        Update the velocity and position of each particle;  }
        
   \Return{the aircraft positions with the optimal fitness}
\end{algorithm}

 \emph{1b) Small-time-scale Trajectory Optimization:}

Regarding the safety-oriented trajectory optimization, a composite potential field (CPF) algorithm is proposed to perform per period $\Delta t$.
The proposed composite potential field includes attractive, repulsive, stable, layered, and goal fields, in which the attractive field is given as,
\begin{equation}
\begin{aligned}
\begin{split}
F_{\text{attr}}=\left\{\begin{array}{c}
\left(d_{i, i_{pre}}-d_{\text {safe }}\right)^2, d_{i, i_{pre}} \geq d_{\text {safe}} \\
0, \ \ \ \ \ \ \ \ \ \ \ \ \ \ \ \ \ \ \ d_{i, i_{pre}}<d_{\text {safe}}
\end{array}\right.
\end{split} 
\end{aligned} 
\label{dcfuvbueyioooo}
\end{equation}
\noindent This attractive field conducts aircraft to chase the route of the preceding aircraft.
In addition, the stable field is, 
\begin{equation}
\begin{aligned}
\begin{split}
F_{\text{stab}}= (v_{i,x} - v_{n_i})^2 + v_{i,y}^2.
\end{split} 
\end{aligned} 
\label{rfewgyueryryeyy} 
\end{equation}
\noindent It primarily pushes aircraft $i$ to maintain specific velocity $v_{n_i}$ in the $n$-th layer to which aircraft $i$ is located.
The repulsive field is written as,
\begin{equation}
\begin{aligned}
F_{\text {repu}}=\left\{\begin{array}{c} 
0, \ \ \ \ \ \ \ \ \ \ \ \ \ \ \ \ d_{i, j} \geq d_{\text {safe}} \\
 \left(\frac{1}{d_{i,j}}-\frac{1}{d_{\text {safe}}}\right)^2, d_{i, j}<d_{\text {safe}}  
\end{array}\right.
\end{aligned} 
\label{gfbh8odsrt_7u8}    
\end{equation}
\noindent, where the repulsive field enforces the distance between adjacent aircraft larger than the safe separation.
Finally, the layer field is given as,
\begin{equation}
\begin{aligned}
\begin{split}
F_{\text {layer}}=\left\{\begin{array}{l} \left(h_i-2 H\right)^2, h_i>\frac{3}{2} H \\
\left(h_i-H\right)^2, \ \frac{1}{2} H<h_i \leq \frac{3}{2} H \\
h_i^2, \ \ \ \ \ \ \ \ \ \ \ h_i \leq \frac{1}{2} H
\end{array}\right.
\end{split} 
\end{aligned} 
\label{dfsiwjjniirerrrwww}
\end{equation}

\noindent It is devised to constrain each aircraft within a discrete layer where $h_i$ is the height of aircraft $i$ and $H$ represents the vertical distance between adjacent layers. Finally, the goal field is expressed as,
\begin{equation}
\begin{aligned}
\begin{split}
F_{\text{goal}}= {||s_{i}(t) - s_{i, goal}^*(t)||}^2,
\end{split} 
\end{aligned} 
\label{sdrgidsrugosrhoeuo}
\end{equation}

\noindent where $s_{i, goal}^*(t)$ is the optimal aircraft position solved by $P4$, which is the outcome of the large-time-scale optimization scheme, aiming to maximize communication rate.  
So the composite potential field for the small-time-scale optimization is,
\begin{equation}
\begin{aligned}
\begin{split}
F_{\text {total}}= w_1^t F_{\text {attr}} + w_2^t F_{\text {stab}} +w_3^t F_{\text {replu}} + w_4^t F_{\text {layer}} +w_5^t F_{\text {goal}}
\end{split} 
\end{aligned} 
\label{sdpoqwiodchgfgf}
\end{equation}

\noindent Meanwhile, we introduce the velocity consensus approach to unify the aircraft velocity in the same layer, which is \cite{drones7040229},
\begin{equation}
\begin{aligned}
a_i=-{\nabla}F_{\text {total}}- \sum\limits_{j=1}^{m_i} \mathcal{I}_{i,j} (v_i-v_j)
\end{aligned} 
\label{dvfsgyeryy}
\end{equation}

\noindent where $\dot{s}_i=v_i, \dot{v}_i=a_i$. $m_i$ is the number of neighbors of aircraft $i$.
And $\mathcal{I}_{i,j}$ is an indicator that takes $1$ when the aircraft $j$ is the neighbor of aircraft $i$, otherwise $\mathcal{I}_{i,j}=0$.

Overall, the RIS-aided trajectory optimization algorithm is shown in Alg.~\ref{Ag2}.
Wherein $\chi$ is the ratio of the large-time-scale period to the small-time-scale period.
$\Delta t$ is the update period of the small-time-scale optimization, and $\chi\Delta t$ is the update period of large-time-scale optimization.

In each small-time-scale period $\Delta t$, the aircraft adjust their acceleration by the composite potential field method for safe aviation. 
In each large-time-scale period $\chi\Delta t$, the RIS communication will be improved by altering the relative positions of low- and high-layer aircraft and RIS phase shift.
Moreover, suppose the traffic density in a layer is too high to maintain the safety separation of adjacent aircraft. It will trigger the layer-switching behavior of aircraft, causing them to switch to a layer with a lower traffic density. The details of the layer-switching scheme will be investigated in the next section.
Finally, each aircraft will update its flight status for better aviation.

{
Generally, our proposed CPF-based small-time-scale trajectory optimization is a distributed algorithm. It can be directly deployed on an individual aircraft. Each potential field can be calculated with a closed-form equation, which means the computational complexity is only proportional to the number of participants.
The computational complexity can be roughly seen as $O(N)$ where $N$ is the number of aircraft involved in the CPF calculation. 
}

\begin{algorithm} 
    \caption{{RIS-aided trajectory optimization}} \label{Ag2}
    
    \textbf{Input}: state of all aircrafts
    
    \textbf{Output}: updated kinetic status of aircraft
    
    \For{$t=\Delta{t}, 2\Delta{t},\cdots\mathcal{T}$}{
        \If{$t\mod \chi\Delta{t}==0$}{
            Calculate the optimal RIS phase shift of low-layer aircraft;\\
            Estimate the optimal position of the low- and high-layer aircraft; 
        }
        \For{i=1,2,3,...,M}{
            \If{Switching layers}{
                Invoking Alg.~\ref{Ag3};
            }
            \Else{
                Calculate aircraft acceleration Eq.~(\ref{dvfsgyeryy});
            }
            
        }
        \For{$i=1,2,3, \cdots M$}{
            Update aircraft kinetic status;
        }
    }			
\end{algorithm}

\subsubsection{ \textbf{Inter-layer Trajectory Optimization}}
\begin{figure}[h]
\centering
     \includegraphics[width=.4\textwidth]{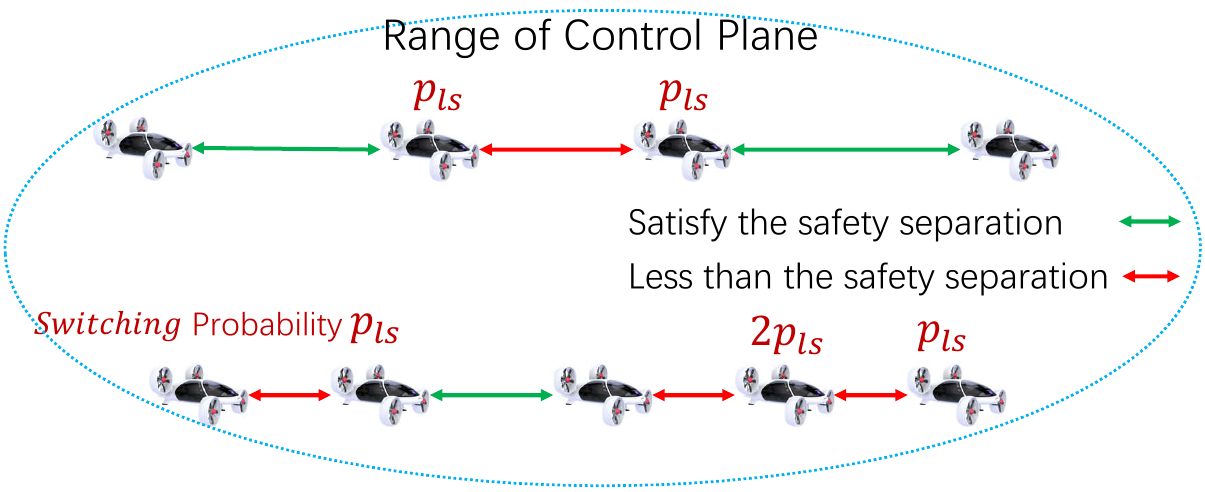} 
     \caption{Layer-switching probability.} 
\label{layer_switchingsss}
\end{figure}

When the flying layer is overcrowded, the horizontal distance between some adjacent aircraft is less than the safety separation.
Suppose an aircraft distance with the preceding or the follower is less than the safety separation. 
In that case, the aircraft will trigger the layer-switching operation with probability $p_{ls}$. 
While the aircraft distances with the preceding $d_{if}$ and the follower $d_{ir}$ are both less than the safety separation, the layer-switching operation is activated with a probability of $2p_{ls}$.
The above procedure is depicted in Fig.~\ref{layer_switchingsss}.

Once an aircraft performs layer-switching operations, it should broadcast a layer-switching request through the omnidirectional channel at the control plane. 
Accordingly, before performing the layer-switching operation, the aircraft monitors the switch tendencies of other aircraft by the omnidirectional channel at the control plane.
Our proposed layer-switching algorithm can ensure collision avoidance of aircraft inspired by the back-off strategy. 

The detailed layer-switching algorithm is shown as Alg.~\ref{Ag3}.
Initially, we set the back-off time $t_r$ to the maximum back-off counts ${t_r}_{max}$.
When an aircraft activates the layer-switching operation, $t_r$ subtracts $1$ in every ${\Delta}t$ period.
If the distance between the adjacent aircraft exceeds the safe separation during the $t_r$ countdown process, the aircraft will cancel the layer-switching operation. Meanwhile, reset $t_r$ to ${t_r}_{max}$.

If the aircraft has detected other layer-switching requests of aircraft in different layers at the control plane, the aircraft must reset the layer-switching operation.
In the reset operation, ${t_r}_{max} = \min\{2 {t_r}_{max}, 32\}$ where $\min\{x,y\}$ represents the minimum value between $x$ and $y$.
The value of ${t_r}$ is selected from $[1, {t_r}_{max}]$ randomly.

As performing the layer-switching process, an aircraft first uniformly accelerates and then decelerates in the vertical direction.
Moreover, the aircraft also needs to uniformly accelerate or decelerate in the horizontal direction to attain the expected horizontal velocity of the target layer. 
When the aircraft is close to the target layer, the composite potential field activates and takes over the flight control for safe aviation in the target layer.

If an aircraft switches from the low layer to the high layer, the acceleration  of the aircraft 
In the layer-switching operation from the low layer to the high layer, the vertical acceleration of an aircraft is calculated as,
\begin{equation}
\begin{aligned}
a_{ls,y}=
\left\{\begin{array}{cc}
    a_{ls,y}^*, & h_i\textless\frac{3}{2}H    \\
    -a_{ls,y}^*, & h_i\geq\frac{3}{2}H
\end{array}\right.,
\end{aligned} 
\label{rewavcrtfdcred}
\end{equation}

\begin{figure}[h]
\centering
     \includegraphics[width=.4\textwidth]{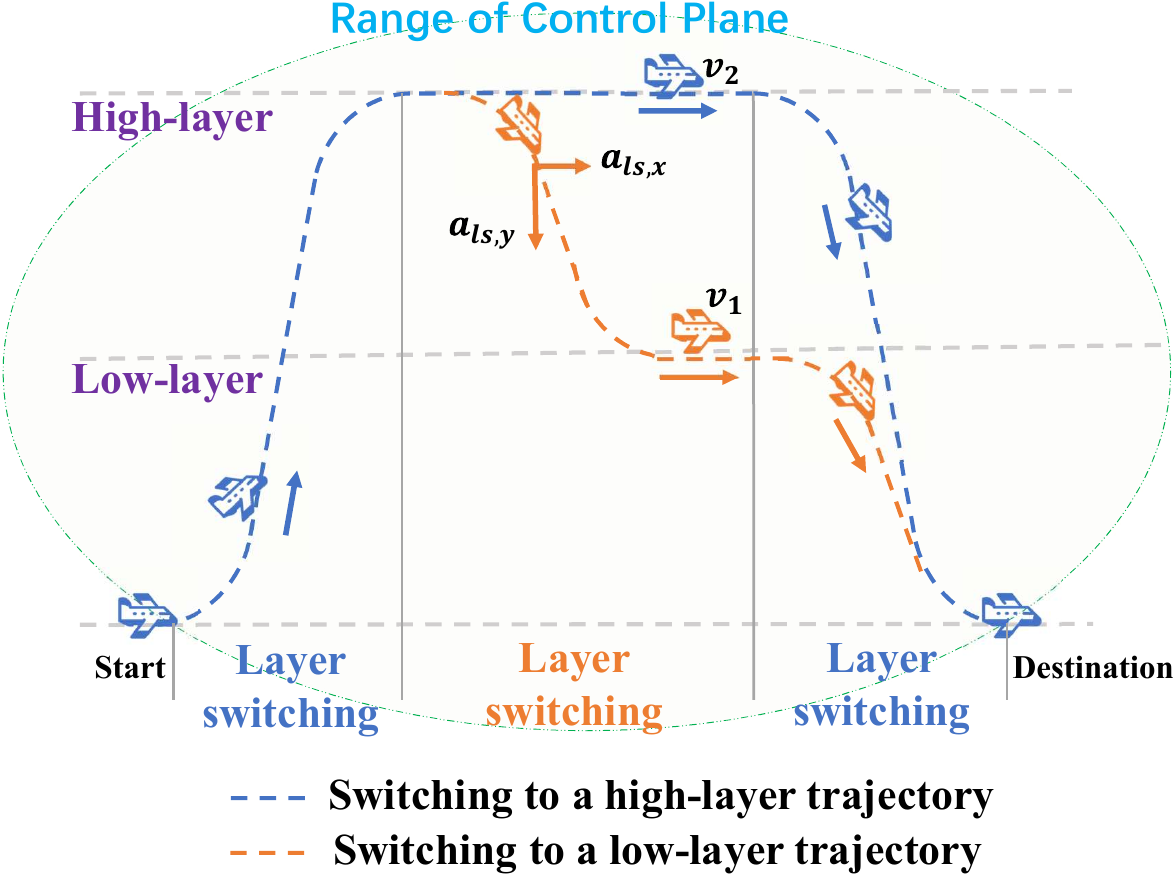} 
     \caption{Demonstration of layer-switching kinetic behaviors.} 
\label{esfuweyufuyuyuyw3ertre}
\end{figure}

\noindent In the layer-switching case, the horizontal velocity must alter from the low-layer average velocity $v_1$ to the high-layer average velocity $v_2$ for smoothly approaching the target layer.
Meanwhile, the vertical velocity should accelerate first and then decelerate. This procedure makes the vertical velocity close to $0$ near the target layer.
To sum up, we obtain the following kinetic equations as,
\begin{equation}
\begin{aligned}
\left\{\begin{array}{cc}
    &v_2-v_1=a_{ls,x}t_{ls}    \\
    &\frac{1}{4}a_{ls,y}^*t_{ls}^2=H     \\
    &a_{ls,x}^2+{a_{ls,y}^{*2}}\le a_{max}^2
\end{array}\right.,
\end{aligned} 
\label{uotruygueeruoefuy}
\end{equation}
\noindent where $t_{ls}$ is the consumption time of the layer-switching. $a_{max}$ is the allowable maximum acceleration. $a_{ls}=[a_{ls,x}, a_{ls,y}]$ is the acceleration of an aircraft, where $a_{ls,y}$ is the vertical acceleration and $a_{ls,x}$ is the horizontal acceleration.
The demonstration of the layer-switching kinetic behaviors is shown in Fig.~\ref{esfuweyufuyuyuyw3ertre}, which demonstrates the average velocity $v_1$ of the high layer, the average velocity $v_2$ of the low layer, the vertical acceleration $a_{ls,y}$ and horizontal acceleration $a_{ls,x}$.

Solving the kinetic equations system Eq.~(\ref{uotruygueeruoefuy}), we get the optimal horizontal acceleration $a_{ls,y}^*$ scheme during layer-switching procedure,
\begin{equation}
\begin{aligned}
a_{ls,y}^*=\frac{\sqrt{{(v_2-v_1)}^4+64H^2a_{max}^2}-{(v_2-v_1)}^2}{8H},
\end{aligned} 
\label{rewavcrtfdcred}
\end{equation}

\noindent where $v_1$ and $v_2$ represent the low- and high-layer average velocities, respectively.
Moreover, according to $a_{ls,y}^*$, the optimal horizontal acceleration $a_{ls,x}$ scheme is given as,

\begin{equation}
\begin{aligned}
a_{ls,x}^*=\frac{(v_2-v_1)\sqrt{a_{ls,y}^*H}}{2H}
\end{aligned} 
\label{rewavcrtfdcred}
\end{equation}

\noindent Based on the optimal acceleration $a_{ls}^*=[a_{ls,x}^*, a_{ls,y}^*]$ and the back-off collision avoidance strategy, we can obtain the safe and efficient layer-switching behavior of aircraft.

\section{Performance Evaluation}

This section conducted numerical simulations of the aforementioned scenarios and algorithms. 
The RIS-aided optimization scheme performs into each time slot $\Delta t$ with discrete phase shifts.
Our proposed algorithm is run on a laptop with AMD Ryzen 7 5800H and coded with Python.
Some important simulation parameters are listed in Tab.~\ref{simulationargs}.

{
\begin{figure*}
\centering
\subfloat[{Dual-plane RIS communication}]{\label{RIS_upper_bound}{\includegraphics[width=0.23\linewidth]{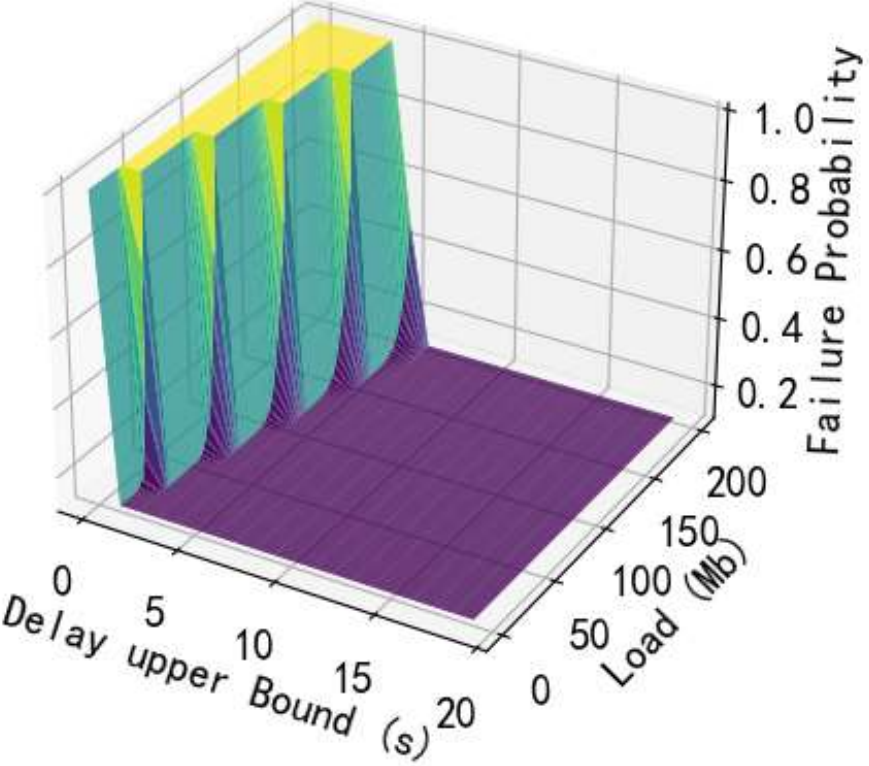}}} 
\subfloat[{Data plane communication}]{\label{Data_plane_upper_bound}{\includegraphics[width=0.23\linewidth]{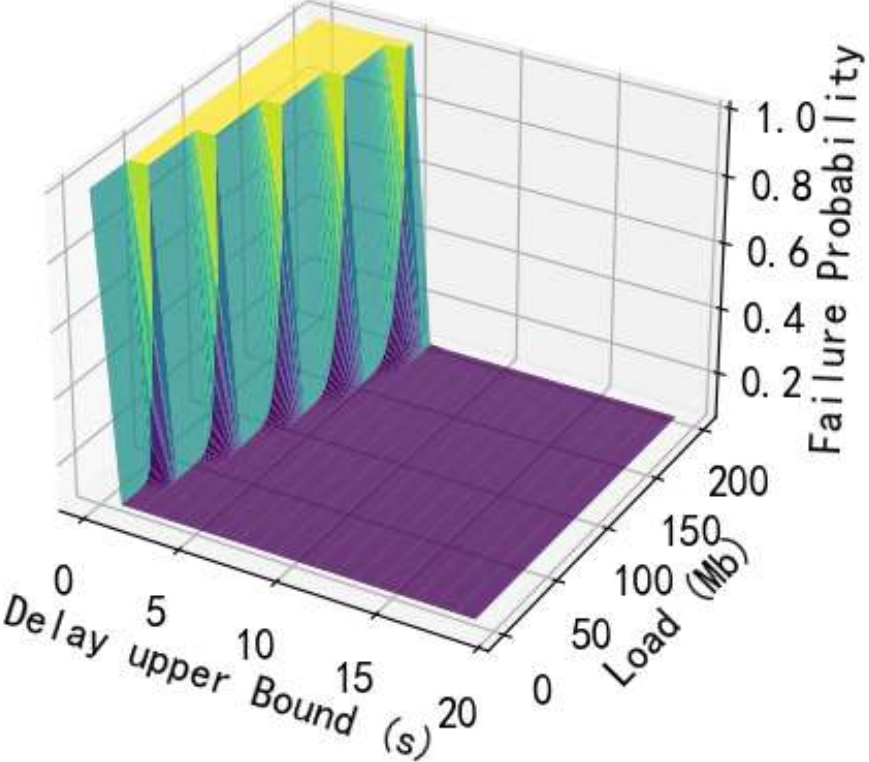}}}
\subfloat[{Control plane communication}]{\label{Control_plane_upper_bound}{\includegraphics[width=0.23\linewidth]{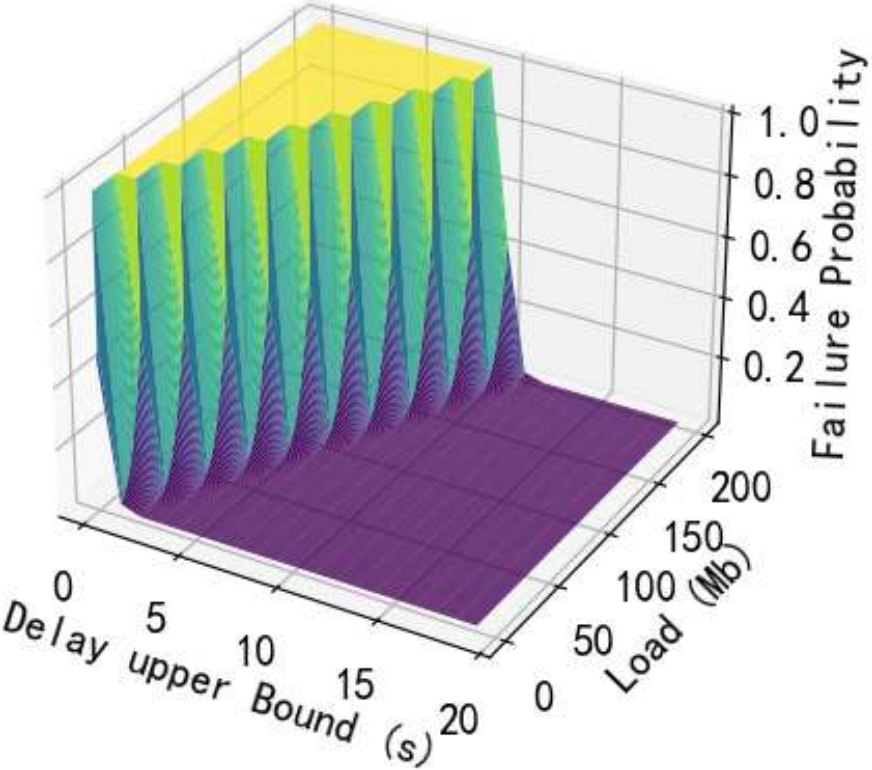}}}
\subfloat[{Probability of delay upper bound beyond 1.5 Seconds.}]{\label{target_delay}{\includegraphics[width=0.25\linewidth]{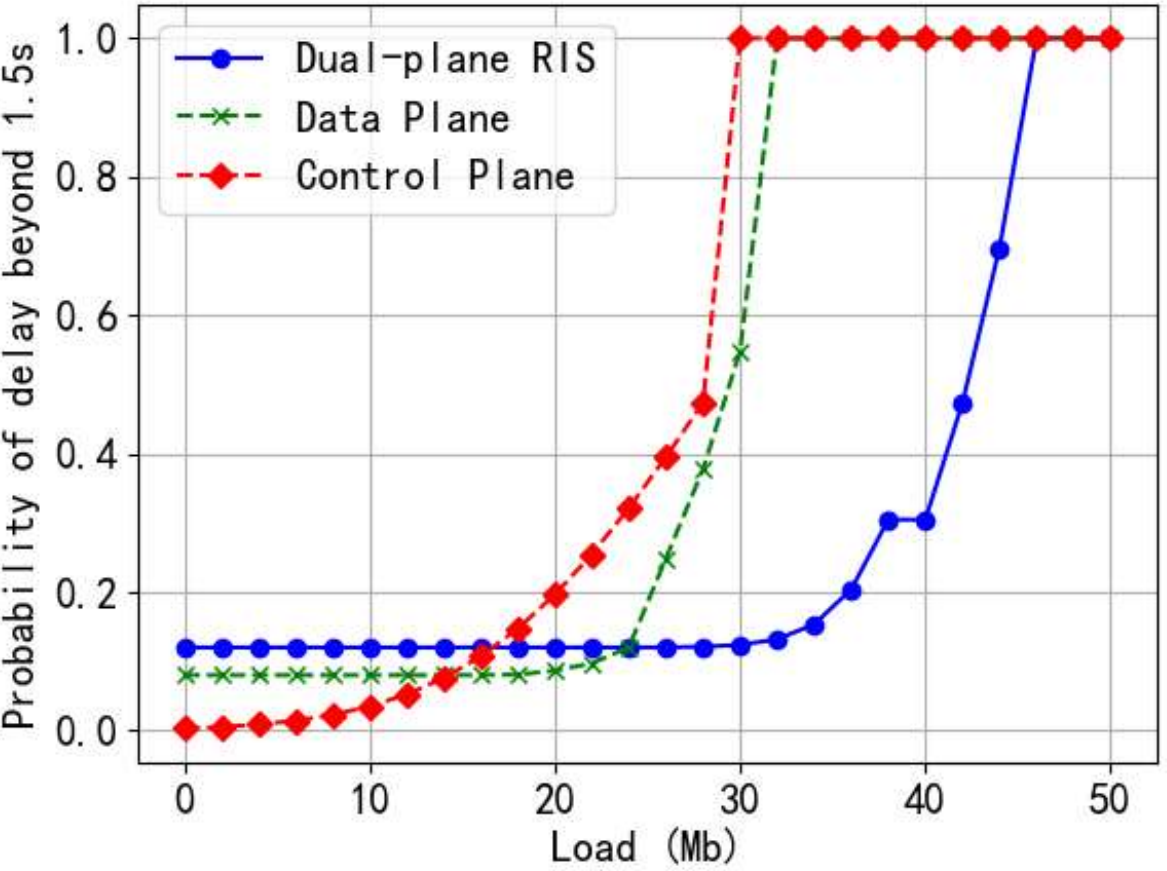}}}
\hfill 
\caption{{Delay upper bound and transmission failure probability vs. network traffic loads.}}
\label{dfasiughoeuwrhgdfsasdPP}
\end{figure*}

The delay upper bounds and transmission failure probabilities of the dual-plane RIS, data plane, and control plane communication are shown in Fig.~\ref{dfasiughoeuwrhgdfsasdPP}.
Wherein the delay upper bound and the transmission failure probability gradually climb with the transmission volume. 
When specifying the communication loads, the transmission failure probability drops as the delay upper bound increases.
The control plane channel shows the worst transmission failure probability and the largest delay upper bound in the three transmission schemes.
In contrast, dual-plane RIS communication has the best performance in terms of delay upper bound and transmission failure probability.

In Fig.~\ref{target_delay}, all three communications demonstrate a trend where the failure probability $P\{D(t)>1.5\}$ (delay bound beyond $1.5s$) increases with the communication loads. 
As the failure probability is $0.2$, the communication load of dual-plane RIS is $40\%$ larger than that of the data plane (direct channel) and $75\%$ larger than that of the control plane.
Moreover, the failure probability of control plane transmission becomes $1$ as the traffic load exceeds $30$ Mb. 
The failure probability of data plane (direct channel) transmission goes to $1$ when the traffic load is over $32$ Mb.
However, the failure probability of dual-RIS transmission reaches $1$ once the traffic load exceeds $45$ Mb. 
This demonstrates that the proposed dual-plane RIS has a higher tolerance for large traffic loads.

It is worth noting that the control plane communication exhibits the lowest failure probability as the traffic load is less than $15$ Mb.
The reason is that aircraft can directly connect to the high-layer aircraft through the omnidirectional channel at the control plane.
A small amount of packets will not cause congestion on the omnidirectional channel.
However, the dual-plane RIS and data plane communications require extra alignment operations, i.e., the RTS, CTS, and RTR processes, which increase transmission latency. 
Thus, the data plane channel prefers light traffic loads. 

Conversely, when the communication loads exceed $24$ Mb, the transmission failure probability $P\{D(t)>1.5\}$ of dual-plane RIS is significantly less than that of others. 
Therefore, employing the dual-plane RIS for large data volume transmission presents a lower delay.
Consequently, dual-plane RIS communication is superior at heavy communication loads, and control-plane communication is sufficient at light loads.
}

\begin{algorithm} 
    \caption{{Layer-Switching Algorithm}} \label{Ag3}
    
    \textbf{Input}: status of all aircrafts, ${t_r}_{max}$, $t_r$
    
    \textbf{Output}: layer-switching operations for aircraft
    
    \textbf{Initialization}: $p_{ls}=0$
    
    \If{$d_{if}\le d_{safe}$ or $d_{ir}\le d_{safe}$}{
        \If{other aircraft have informed layer-switching requests}{
            Perform random back-off;
        }
        \If{$d_{if}\le d_{safe}$ and $d_{ir}\le d_{safe}$}{
            Switching the layer with probability $p_{ls}$}
        \Else{
            Switching the layer with probability $2p_{ls}$
        }
        
    }  
    \Else{
        Maintain the current aircraft status
    }			
\end{algorithm}

%

 \begin{table}[!ht]
    \centering
    \caption{Simulation Parameters}
    \begin{tabular}{|l|l|l|}
    \hline
        \textbf{Description} & \textbf{Value} \\ \hline
       Rate of RTS/CTS/RTR Transmission & 20 \\ \hline
       Data Volume of RTS/CTS/RTR Message & 3 \\ \hline
       Rate of RIS Channel & 80-100 \\ \hline
       Rate of Direct Channel & 40 \\ \hline
       Rate of Omnidirectional Channel & 20 \\ \hline
       Reference Channel Coefficient $\beta$ & -30dB \\ \hline
       Path Loss between BS and Low-layer Aircraft $k$ ${\alpha}_{BS,k}$ & 2.5 \\ \hline
       Path Loss between BS and High-layer Aircraft $i$ ${\alpha}_{BS,i}$ & 2 \\ \hline
       Path Loss between Aircraft $i$ and $k$ ${\alpha}_{i,k}$ & 2.2 \\ \hline
       Transmission Power of BS $P_{BS}$ & 300dBm \\ \hline
       Noise Power ${\sigma}^2$ & -169dBm \\ \hline
       Simulation Time Step ${\Delta}t$ & 0.1s \\ \hline
       Expected Velocity of the Aircraft on the Ground $v_0$ & 30m/s \\ \hline
       Expected Velocity of the Low-layer Aircraft $v_1$ & 45m/s \\ \hline
       Expected Velocity of the High-layer Aircraft $v_2$ & 60m/s \\ \hline
       Large-time-scale Multiple q & 5 \\ \hline
       Layer Switching Probability $p_{ls}$ & 0.4 \\ \hline
       Interference Power & 1dBm \\ \hline
       Interference Position & (800,100) \\ \hline
       Stationary RIS Position & (400,100) \\ \hline
       
    \end{tabular}
    \label{simulationargs}
\end{table}

\begin{figure*}
\centering
\subfloat[RIS mounted on the low-layer aircraft]{\label{Airborne_RIS}{\includegraphics[width=0.32\linewidth]{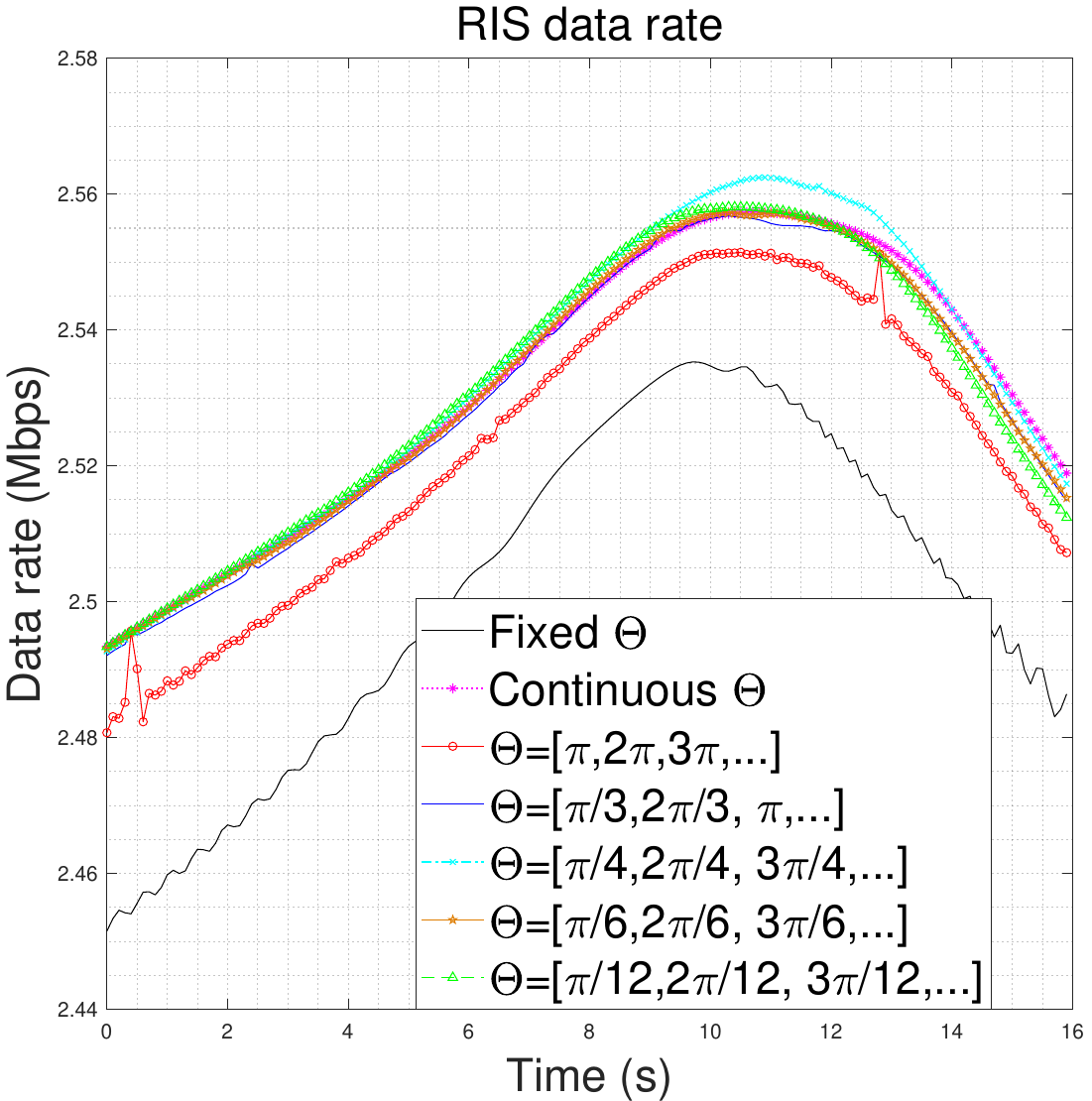}}} 
\subfloat[Airborne RIS with interference]{\label{Airborne_RIS_int}{\includegraphics[width=0.32\linewidth]{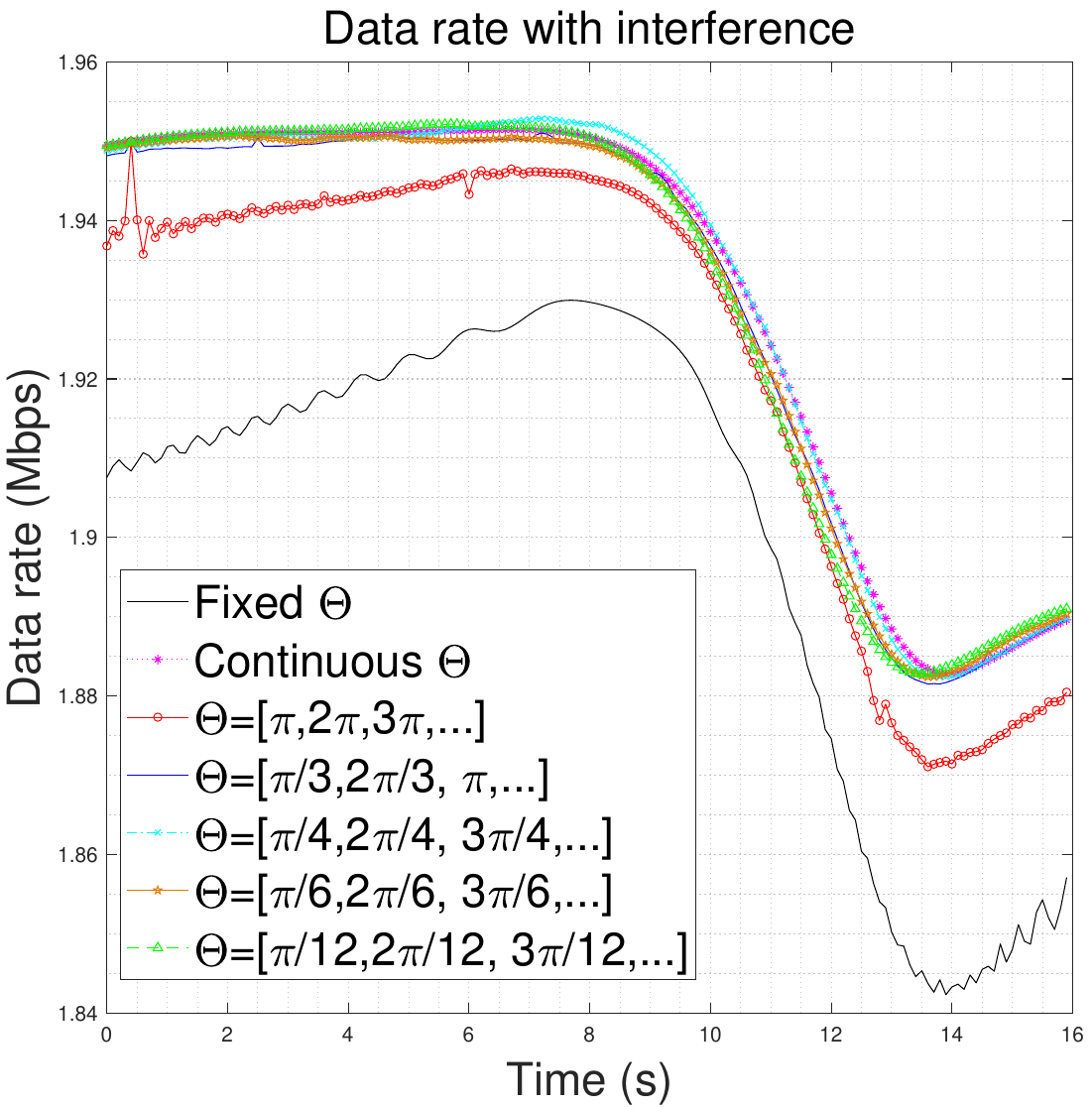}}}
\subfloat[Stationary RIS of building surfaces]{\label{Stationary_RIS}{\includegraphics[width=0.32\linewidth]{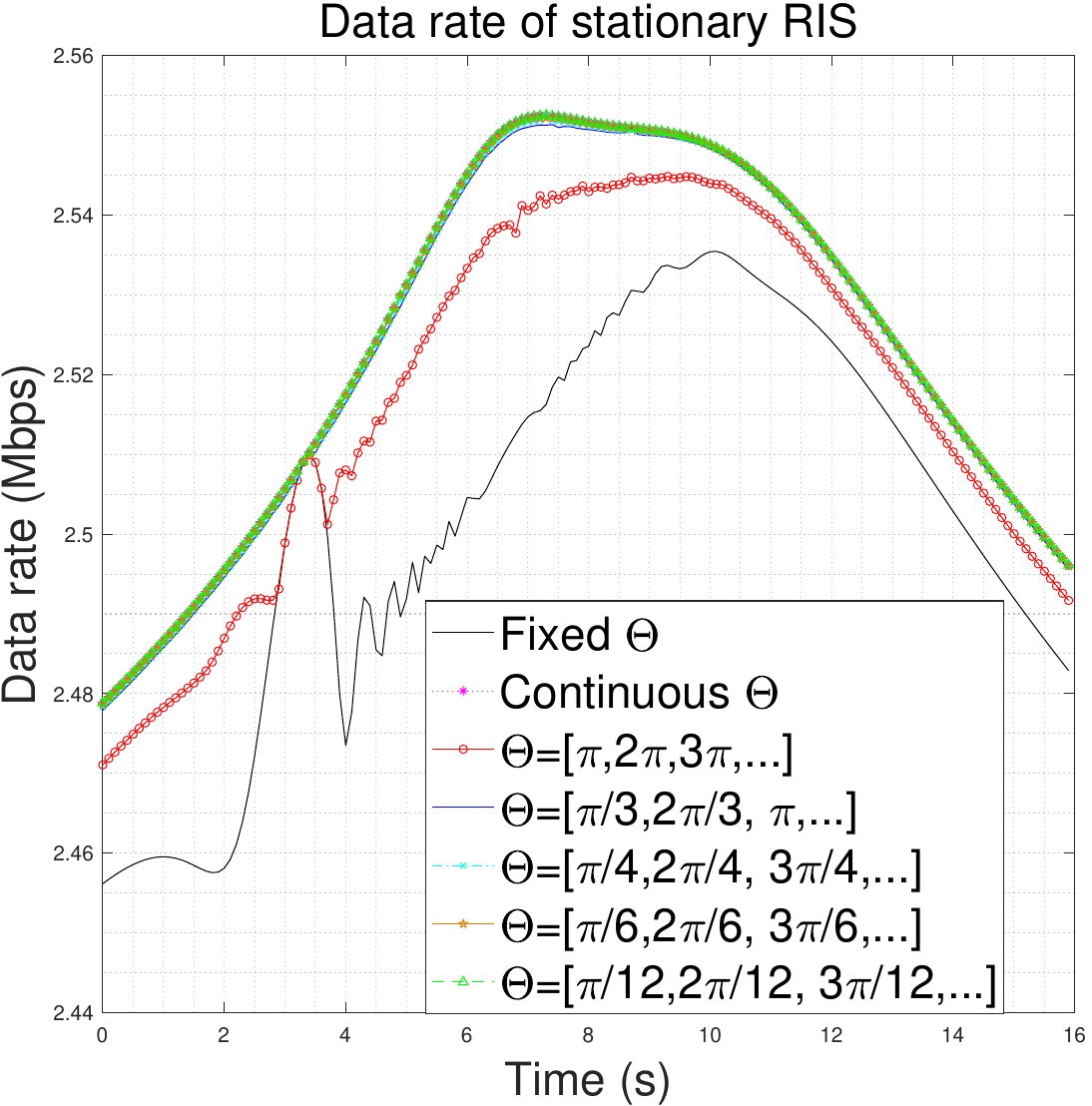}}}
\hfill 
\caption{Data rate from the BS to high-layer aircraft during navigation.}
\label{fugyodzweigsiu}
\end{figure*}

\begin{figure*}
\centering
\subfloat[Trajectory of aircraft]{\label{trajectory_Airborne_RIS}{\includegraphics[width=0.33\linewidth]{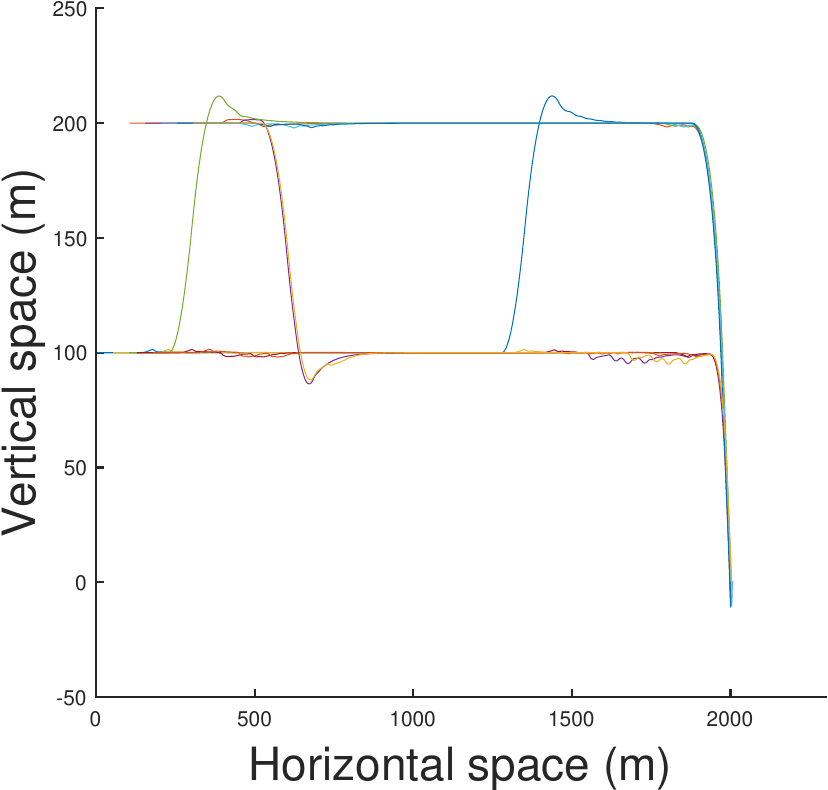}}} 
\subfloat[Trajectory of aircraft with interference]{\label{trajectory_Airborne_RIS_int}{\includegraphics[width=0.33\linewidth]{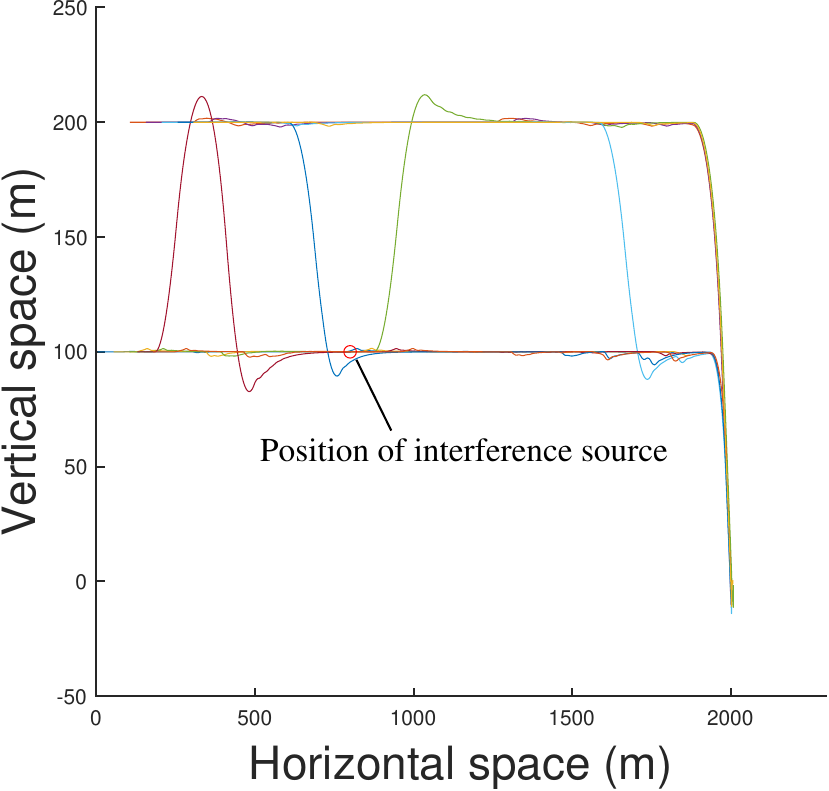}}}
\subfloat[Trajectory of aircraft with stationary RIS]{\label{trajectory_Stationary_RIS}{\includegraphics[width=0.33\linewidth]{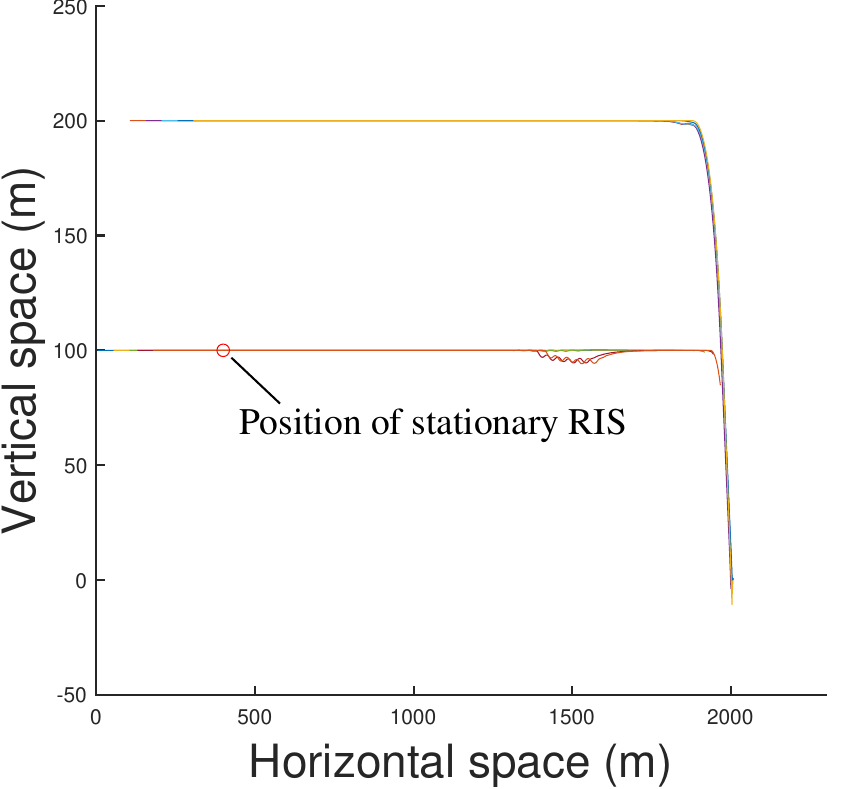}}}
\hfill 
\caption{Aircraft trajectories are affected by different communication configurations.}
\label{rguyuygvgygyuguyt}
\end{figure*}

Fig.~\ref{fugyodzweigsiu} demonstrates the data rate from the BS to the high-layer aircraft in various RIS configurations.
Fig.~\ref{Airborne_RIS} shows the data rate of the RIS mounted on the low-layer aircraft during aviation.
In this scenario, there are several RIS phase shift control schemes, i.e., the fixed phase $\Theta$, the continuous phase shift, the discrete phase shift with a resolution of ${\pi}$, $\frac{\pi}{3}$, $\frac{\pi}{3}$, $\frac{\pi}{4}$, $\frac{\pi}{6}$, and $\frac{\pi}{12}$.
Due to manufacturing limitations, the practical RIS phase shift can only take discrete values.
In addition, the continuous phase shift does not need the large-time-scale trajectory scheme since the continuous shift can achieve the optimal transmission rate for any position of the high-layer aircraft according to Eq.~(\ref{dfgsuhryueueryry}).
As the benchmark, the continuous phase shift scheme has the optimal transmission rate during all aviation.

In Fig.~\ref{Airborne_RIS}, the transmission rate of these discrete phase shift schemes is close, outperforming the fixed scheme $\Theta =\bf{0}$.
As the phase shift resolution is less than $\frac{\pi}{3}$, the transmission rate curves of the discrete phase shift have a rough same trend line as the continuous phase shift curve.

Fig.~\ref{Airborne_RIS_int} illustrates the RIS transmission rate from the BS to the high-layer aircraft with interference where the interference source position is $(800, 100)$ and the interference power is $1$ dBm.
The overall transmission rate with interference is lower than the rate without interference. 
During $[10s, 12s]$ interval, the tranmission rate from the BS to the high-layer aircraft experiences significant attenuation in that the low-layer aircraft with RIS closes the interference source.
Similarly, while the phase shift resolution is less than $\frac{\pi}{3}$, the discrete phase shift curve is proximate to the continuous phase shift curve.

Fig.~\ref{Airborne_RIS_int} gives the transmission rate of a stationary RIS.
The stationary RIS can be deployed on a building surface rather than the moved low-layer aircraft.
Regarding the signal range of BSs, the altitude of the stationary RIS is identical to the low layer, which is also sound for the performance comparisons.
As described in Fig.~\ref{Stationary_RIS}, the data rate of fixed phase shift intense changes over time since this stationary RIS with fixed phase shift has few degrees of freedom to optimize RIS transmission.
Moreover, the discrete phase shift curve is similar to the continuous curve when the phase shift resolution is less than $\frac{\pi}{3}$.
\begin{figure}
\centering
\subfloat[Horizontal velocity]{\label{Horizontal_Velocity}{\includegraphics[width=0.49\linewidth]{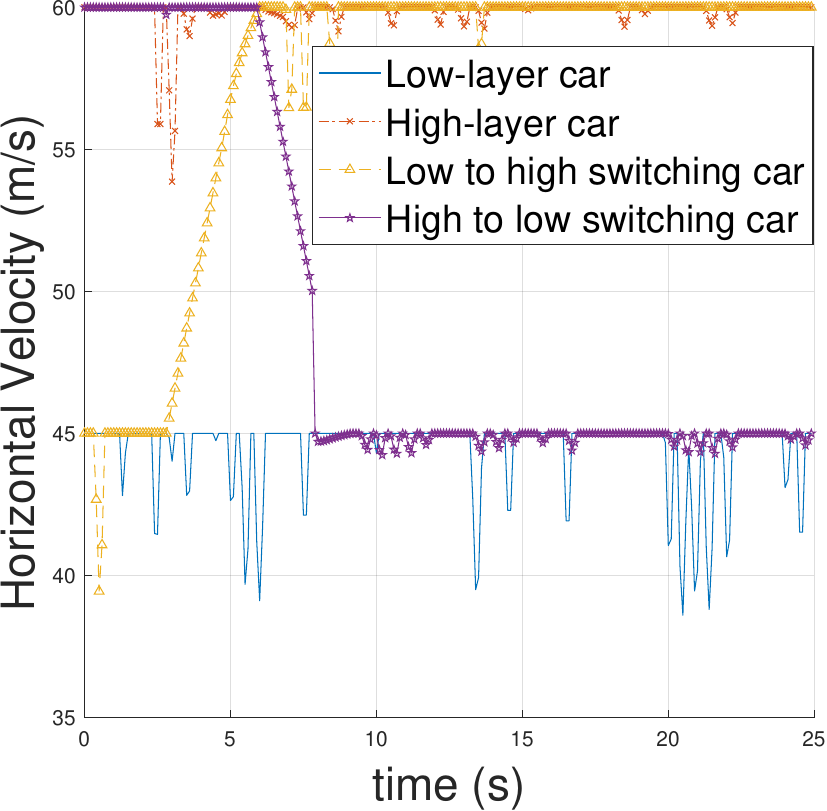}}} 
\subfloat[Vertical velocity]{\label{Vertical_Velocity}{\includegraphics[width=0.5\linewidth]{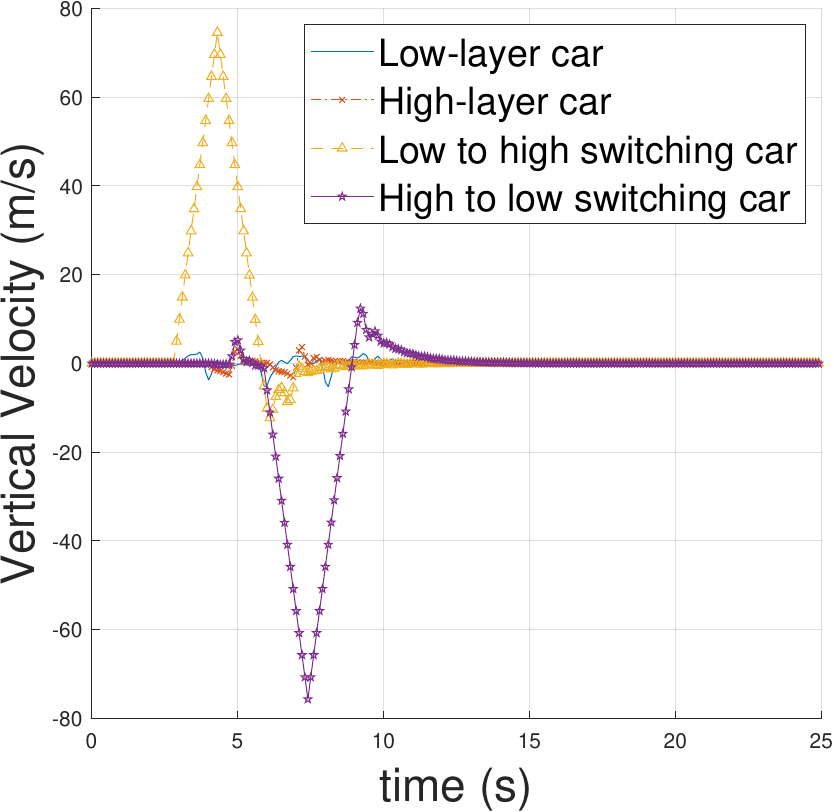}}}
\hfill 
\caption{Velocity of the horizontal and vertical aviation in layered UAM.}
\label{sdvidfujvsuifdhuiho}
\end{figure}

Subsequently, we investigate how the communication configurations affect the aircraft trajectories. 
In Fig.~\ref{rguyuygvgygyuguyt}, there are two airways, i.e., low- and high-layer airways.
The heights of the low layer and high layer are $100m$ and $200m$, respectively. 
To proceed, we place $5$ aircraft in each layer based on our proposed scheme to aviate during the range of $[0m, 2000m]$.

Fig.~\ref{trajectory_Airborne_RIS} is the aircraft trajectories with the airborne RIS mounted on low-layer aircraft.
Few aircraft perform layer-switching procedures triggered by violating safety separation.
In Fig.~\ref{trajectory_Airborne_RIS_int}, 
we observe that the number of layer-switching operations increases.
The reason is that the communication quality determines the trajectory aviation by the proposed small-time-scale trajectory scheme.
Thus, the interference renders erratic safety separations by deteriorating communication quality.
The erratic safety separation is the main reason for frequent layer-switching.

\begin{figure}
\centering
\subfloat[Composite potential field]{\label{total_apf}{\includegraphics[width=0.33\linewidth]{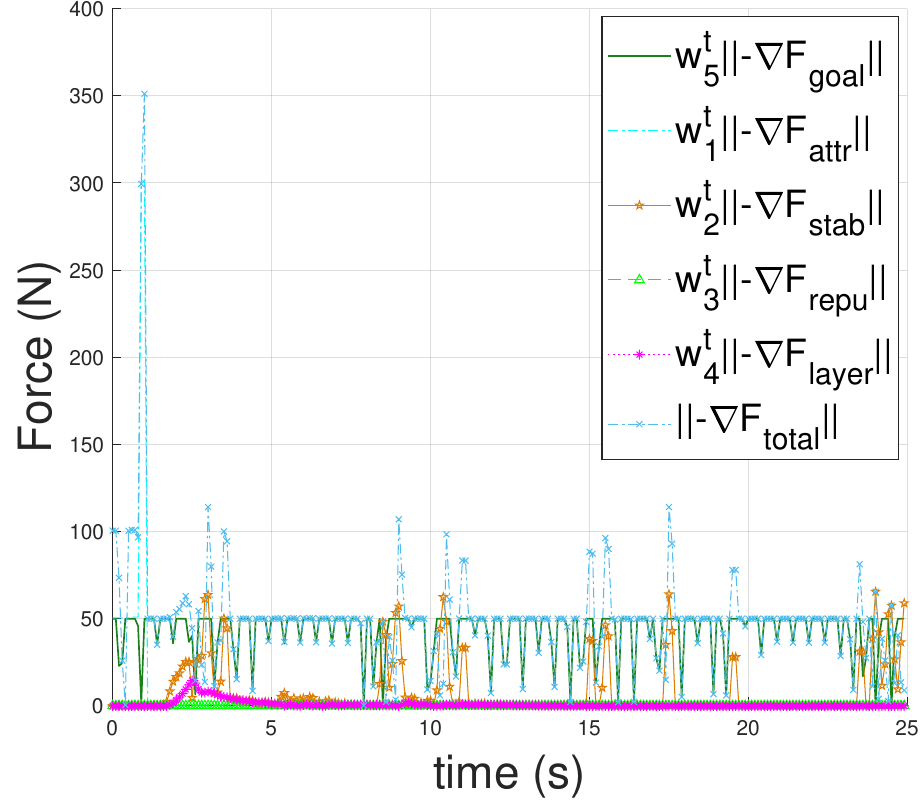}}} 
\subfloat[Composite potential field without the goal field]{\label{total_apf_without_goal}{\includegraphics[width=0.33\linewidth]{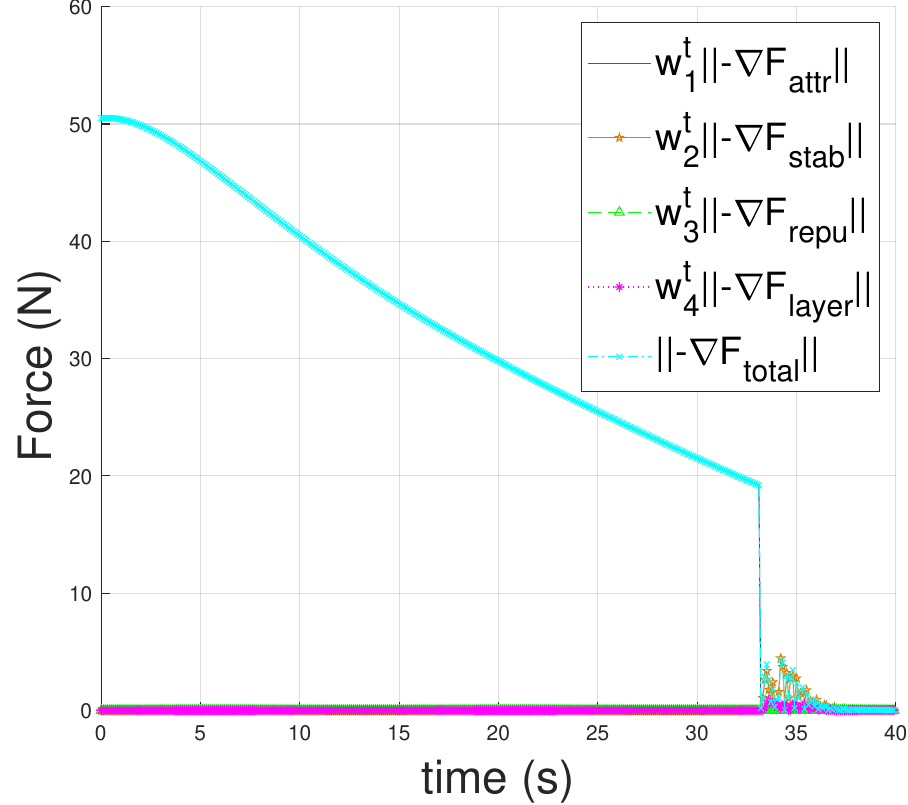}}}
\subfloat[Repulsive field]{\label{Repu_field}{\includegraphics[width=0.33\linewidth]{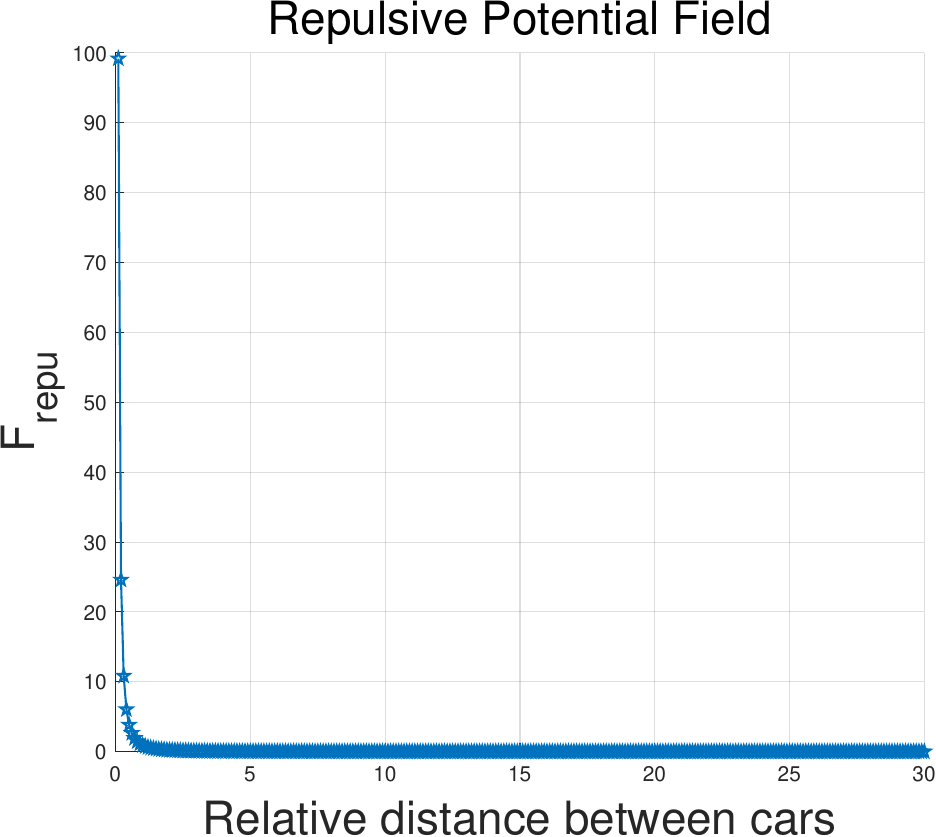}}} 
\hfill 
\subfloat[Attractive field]{\label{Attr_field}{\includegraphics[width=0.33\linewidth]{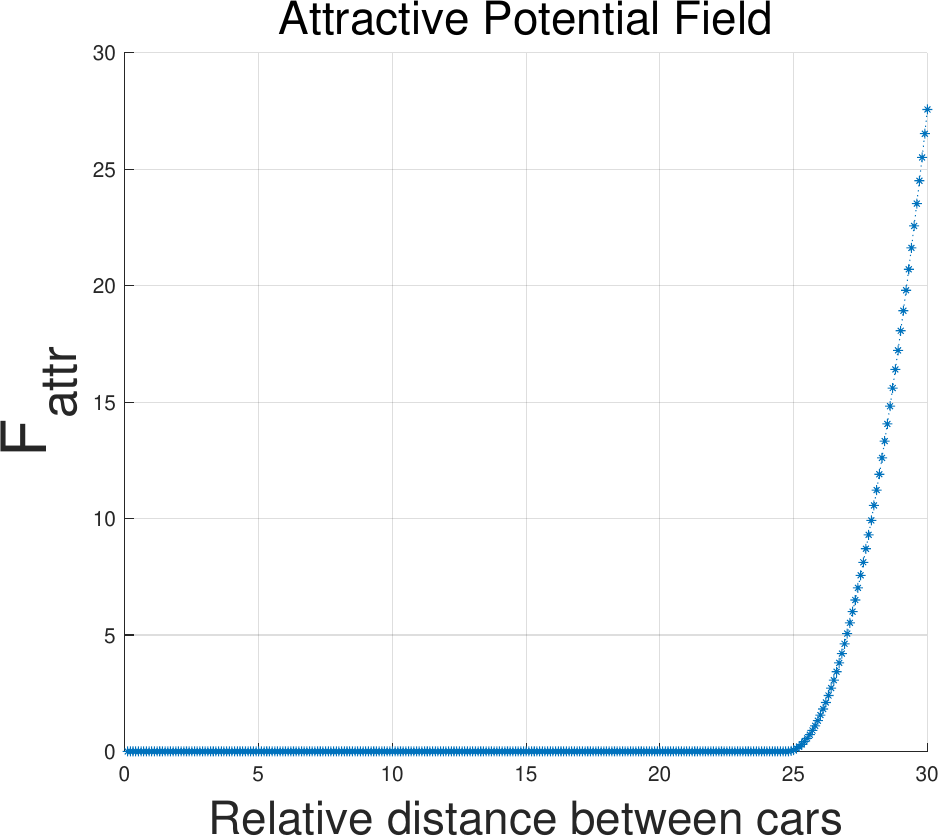}}}
\subfloat[Stable field]{\label{stable_field}{\includegraphics[width=0.33\linewidth]{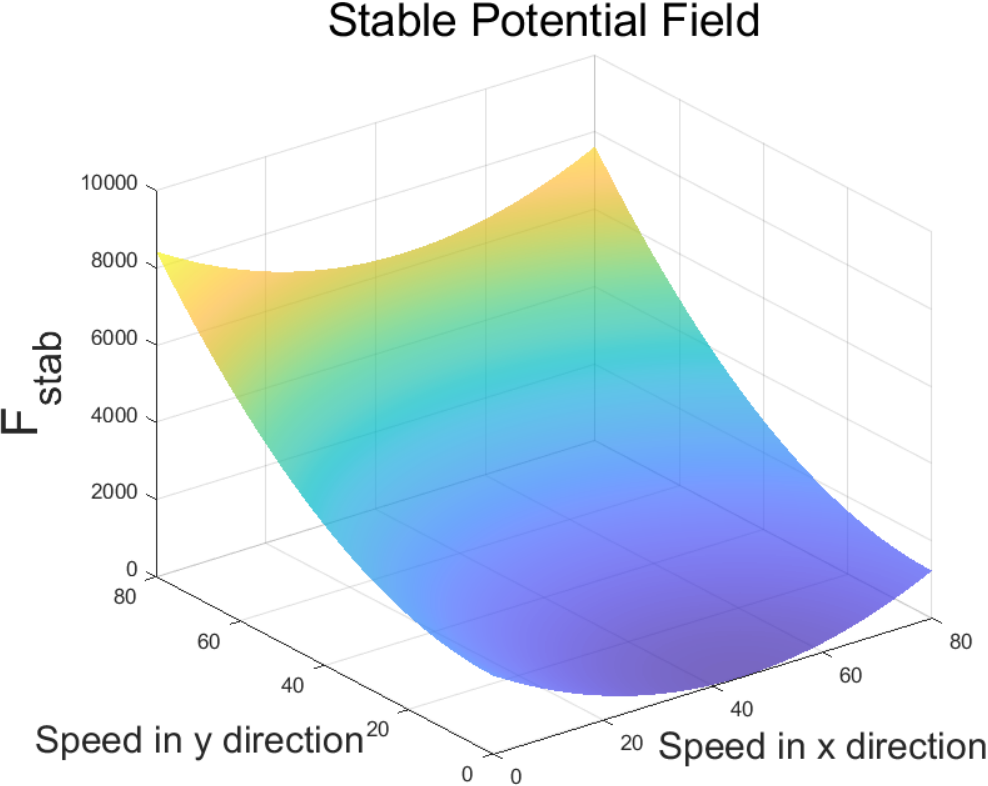}}} 
\subfloat[Layered field]{\label{Layer_field}{\includegraphics[width=0.33\linewidth]{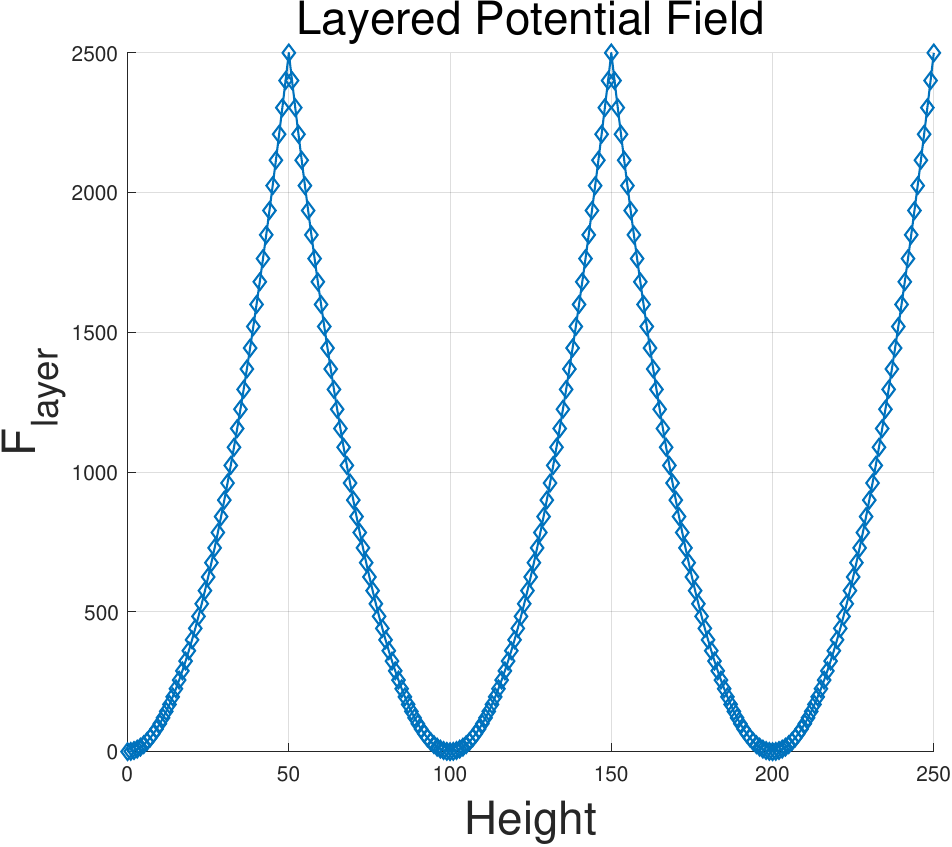}}}
\hfill 
\caption{Variations of the composite potential fields over time.}
\label{dsukueueruei}
\end{figure}

Fig.~\ref{trajectory_Stationary_RIS} is the aircraft trajectory with a stationary RIS deployed on a building surface. 
A stationary RIS conveys a relatively stable communication condition, preventing significant fluctuations in large-scale trajectory optimization. 
In addition, the proposed small-time-scale trajectory scheme tries to maintain a proper safety separation during aircraft aviation.
Without the fine-tuning of the large-time-scale trajectory scheme, aircraft will not violate the safety separation, leading to no layer-switching of aircraft.

To comprehensively display the aircraft aviation process, Fig.~\ref{sdvidfujvsuifdhuiho} compares velocity variation over time in horizontal and vertical directions.
In Fig.~\ref{Horizontal_Velocity}, the velocities of low- and high-layer aircraft are roughly $45 \ m/s$ and $60 \ m/s$, respectively.
During $3s$ to $7s$, some aircraft switch from the low to the high layer and from the high to the low layer.
Due to the velocity difference between the low layer and high layer, the switched low-layer aircraft needs to accelerate to cater to the reference speed of the high layer.
In contrast, the switched high-layer aircraft needs to decelerate to cater to the reference speed of the low-layer.
The acceleration and deceleration of the layer-switching are both uniformly accelerated motions.
Compared to the horizontal velocity, the vertical velocities of aircraft are close to zero.
This is because our proposed scheme can effectively stabilize the aircraft distribution in the layered airspace structure.

As illustrated in Fig.~\ref{dsukueueruei}, the variation of the composite potential field over time is analyzed according to different potential field equations Eq.~(\ref{dcfuvbueyioooo})-(\ref{sdrgidsrugosrhoeuo}).
The whole composite potential fields are shown in Fig.~\ref{total_apf} where $\Delta F_{total}$ represents the joined force of the composite fields. 
Within $2s$, the joined force fluctuates violently. Beyond $2s$, the joined force approaches to a stable value.
From Fig~\ref{total_apf_without_goal}, we can see that this value is determined by the goal potential field since the joined force without the goal field converges to $0$ over time.
The repulsive, attractive, stable, and layered potential field kinetic graphs with time are shown in Fig.~\ref{Repu_field}, Fig.~\ref{Attr_field}, Fig.~\ref{stable_field}, and Fig.~\ref{Layer_field}, respectively.

\begin{figure}
\centering
\subfloat[Each layer has 5 aircraft]{\label{Horizontal_Velocity}{\includegraphics[width=0.5\linewidth]{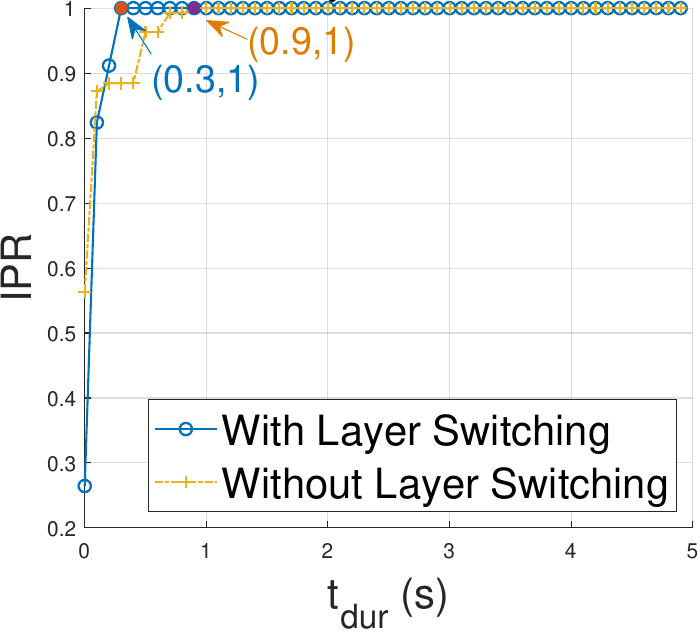}}} 
\subfloat[Each layer has 20 aircraft]{\label{Vertical_Velocity}{\includegraphics[width=0.5\linewidth]{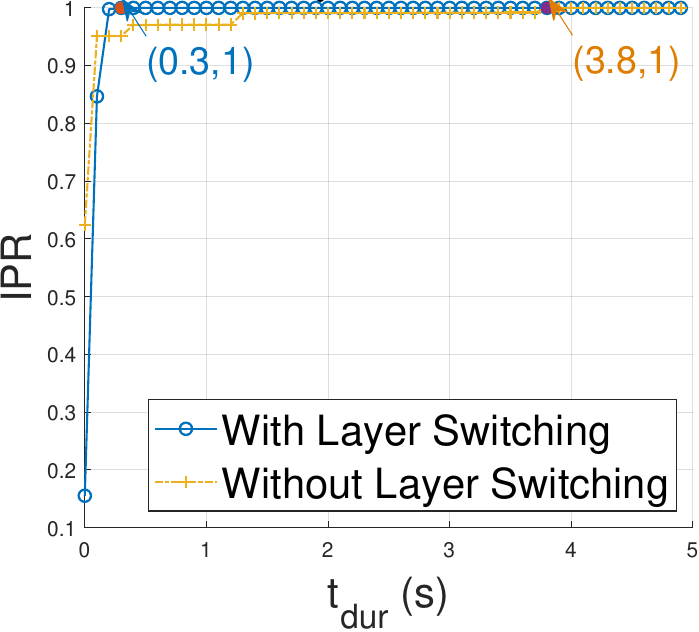}}}
\hfill 
\subfloat[Each layer has 30 aircraft]{\label{Repu_field}{\includegraphics[width=0.5\linewidth]{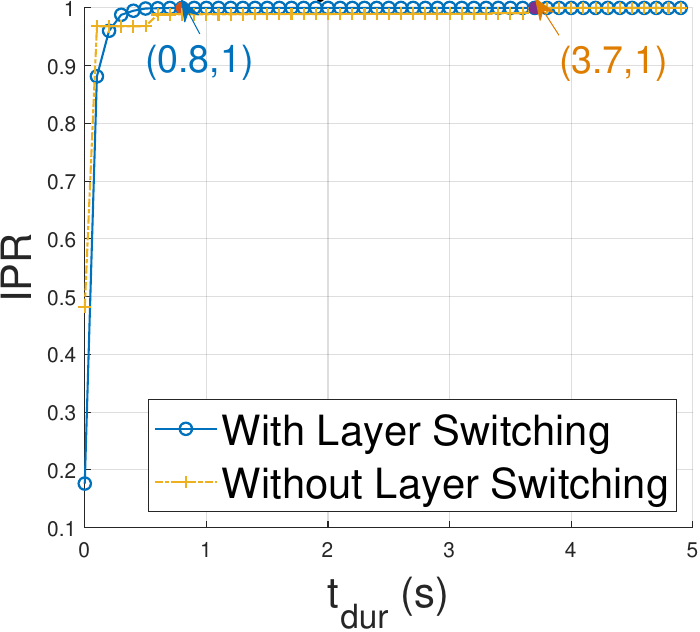}}} 
\subfloat[Each layer has 50 aircraft]{\label{Attr_field}{\includegraphics[width=0.5\linewidth]{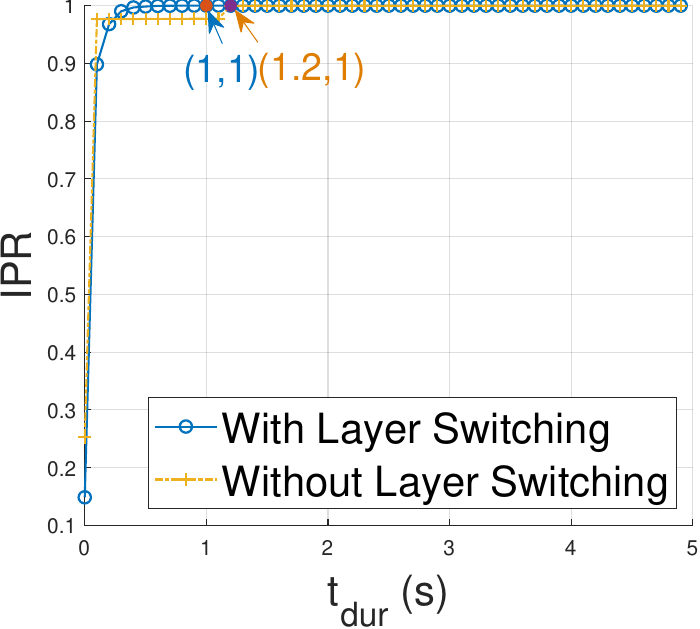}}}
\hfill 
\caption{IPR vs. duration threshold $t_{dur}$.}
\label{dsfguyihertugearuydfvguydsaz}
\end{figure}

To measure safe aviation performance, we investigate the intrusion prevention rate (IPR) metric with different numbers of aircraft.
As shown in \cite{42358726Sunil}, this metric describes the proportion of valid safe separation between adjacent aircraft, which is given as,
\begin{equation}
\begin{aligned}
\begin{split}
IPR= \frac{n_{cfl}-n_{int}(t_{dur})}{n_{cfl}},
\end{split} 
\end{aligned} 
\label{fdjlairenwlakdj}
\end{equation}

\noindent where $n_{cfl}$ represents the number of conflicts. A conflict is defined as the distance between adjacent aircraft being less than the safety separation. 
$n_{int}(t_{dur})$ represents the number of intrusions with threshold $t_{dur}$. 
If the adjacent distance is less than the safe separation and the duration exceeds $t_{dur}$, we regard it as an intrusion with threshold $t_{dur}$.
Typically, IPR increases with $t_{dur}$ and comes to $1$ once $t_{dur}$ surpasses a threshold.
 
Fig.~\ref{dsfguyihertugearuydfvguydsaz} performs the IPR with duration thresholds $t_{dur}$ of different number of aircraft.
It illustrates the specific duration threshold $t_{dur_{(IPR=1)}}$, making IPR attain $1$.
We can see that $t_{dur_{(IPR=1)}} = 0.3$ as the aircraft number of each layer is $5$. Compared with the non-layer switching, the time of restoring aircraft to a safe separation is shrunk by $66\%$. 
Fig.~\ref{dsfguyihertugearuydfvguydsaz} also indicates that the airline has not yet reached saturation when the aircraft number of each layer is below $20$.
However, if the aircraft number of each layer exceeds $30$, $t_{dur_{(IPR=1)}} \ge 0.8$.
More time is consumed to restore the distance between adjacent aircraft to the safe separation since the layered airline may have reached saturation as the aircraft number exceeds $30$.
In addition, Fig.~\ref{dsfguyihertugearuydfvguydsaz} also reveals that the layer switching scheme can significantly reduce $t_{dur_{(IPR=1)}}$ at different numbers of each layer aircraft. 
The reason is that the layer-switching operation allows each aircraft to pick its proper layer. It can adaptively adjust the number of aircraft in each layer to avoid the situation of congestion in one layer while adjacent layers are idle.



{
\section{Conclusion}

This paper proposed a RIS-aided trajectory optimization scheme for safe aviation and efficient communication in layered UAM. 
The proposed dual-plane RIS communication has the lowest delay upper bound compared to other benchmarks.
According to this communication configuration, we propose a dual-time-scale scheme to optimize the aircraft trajectory in the horizontal dimension.
It comprises large- and small-time-scale optimizations.
The large-time-scale optimization adopted the particle swarm method to acquire the positions with the maximum transmission rate.
The small-time-scale optimization applied the composite potential field method to ensure safe and stable aircraft flight in the layered UAM.
In addition, we develop a layer-switching method to conduct safe inter-layer aviation in the vertical dimension.
This work provisions the possibility of utilizing 6G communication technologies in the trajectory optimization of layered airspace.

Note that the proposed joint optimization scheme is designed for the layered airspace with parallel flight corridors, regarding the intra-layer (horizontal) and inter-layer (vertical) flight controls. However, the practical airspace may not be just a set of parallel corridors. There could be a complex 3-dimensional spatial structure, such as several spiral lines.
In future work, we will expand the proposed flight control scheme within diverse 3-dimensional structured airspace.

}

\appendices
\footnotesize
\bibliography{biblio}
\end{document}